\documentclass[reprint,aps,pre,amsmath,amssymb,unsortedaddress]{revtex4-2}
\usepackage{graphicx} % Required for inserting images
\usepackage{subcaption}
\usepackage{amsmath,amsthm,amsfonts,bbm,bm}
\usepackage{hyperref}
\theoremstyle{plain}
\newtheorem{thm}{Theorem}

\theoremstyle{definition}

\newtheorem{exmp}[thm]{Example}
\theoremstyle{remark}
\newtheorem{rem}[thm]{Remark}

\newcommand{\One}{\mathbbm{1}}
\newcommand{\NN}{\mathbbm{N}}
\newcommand{\QQ}{\mathbbm{Q}}
\newcommand{\RR}{\mathbbm{R}}
\newcommand{\ZZ}{\mathbbm{Z}}

\begin{document}

\title{Machine Learning Conservation Laws of Dynamical systems}
\author{Meskerem Abebaw Mebratie}
\affiliation{Institut f\"ur Mathematik, Universit\"at Kassel, 34109 Kassel}
\author{R\"udiger Nather}
\affiliation{Fachbereich Elektrotechnik und Informatik, Universit\"at Kassel, 34121 Kassel}
\author{Guido Falk von Rudorff}
\affiliation{Institut f\"ur Chemie, Universit\"at Kassel, 34109 Kassel}
\affiliation{Center for Interdisciplinary Nanostructure Science and
  Technology (CINSaT), Heinrich-Plett-Stra\ss e 40, 34132 Kassel} 
\author{Werner M. Seiler} 
\affiliation{Institut f\"ur Mathematik, Universit\"at Kassel, 34109 Kassel}
\date{\today}

\begin{abstract}
  Conservation laws are of great theoretical and practical interest.  We
  describe a novel approach to machine learning conservation laws of
  finite-dimensional dynamical systems using trajectory data.  It is the
  first such approach based on kernel methods instead of neural networks
  which leads to lower computational costs and requires a lower amount of
  training data.  We propose the use of an ``indeterminate'' form of kernel
  ridge regression where the labels still have to be found by additional
  conditions.  We use here a simple approach minimising the length of the
  coefficient vector to discover a single conservation law.
\end{abstract}

\maketitle

\section{Introduction}

Conservation laws of dynamical systems are of great theoretical and
practical interest.  In physics, many fundamental principles take the form
of a conservation law with the conservation of energy probably the most
prominent example.  But also in biological and chemical models,
conservation of mass and other conservation principles play a prominent
role.  On a more theoretical side, knowledge of a conservation law almost
always provides important insight into a system.  Practically, conservation
laws may e.g.\ allow for a model reduction, if certain degrees of freedom
can be expressed through others and thus be eliminated from the model
equations.

Recently, there have been some efforts in discovering conservation laws of
dynamical systems via machine learning, see e.g.\
\cite{kkb:dcl,wmspg:siam,hj:dcl,lt:poin1,lmt:poin2,lsbst:new,abbh:conslaw,lds:clotml,zzk:mlclnd}
and references therein.  In these works, quite different characterisations
of conservation laws and quite different techniques from machine learning
are used.  Some approaches identify only a single conservation law, others
try to find all of them.  Most approaches are based on trajectory data,
i.\,e.\ they do not require knowledge of the dynamical system
(\ref{eq:dynsys}) itself --- they are ``model-agnostic'', as the authors of
\cite{abbh:conslaw} call it ---, but \cite{lmt:poin2} uses explicitly the
dynamical system and no trajectory data.  Common to all these works is the
use of neural networks as basic technology and that only well-known
textbook examples of conservation laws have been ``discovered''.  A
conservation law is usually also found as a neural network approximating
it.  Most of the cited works then try in a subsequent symbolic regression
step to obtain a closed-form expression for the conservation law.  Judging
from the presented examples, this seems to work pretty well for
conservation laws which are essentially rational functions, but
difficulties arise when transcendental functions with non-trivial arguments
appear.

In this article, we present a novel approach to machine learning
conservation laws using \emph{kernel methods}, more precisely \emph{kernel
  ridge regression} \cite{syk:kernel,ss:kernel,stc:kernel}.  On the
downside, this implies that we can only discover conservation laws living
in the Hilbert space defined by the used kernel.  Our basic tool is the
inhomogeneous polynomial kernel meaning that we search mainly for
polynomial conservation laws, but we will discuss how this restriction can
be lifted.  On the upside, kernel methods always provide explicit
closed-form expressions for the discovered conservation laws so that no
subsequent symbolic regression is necessary.  We furthermore believe that
such an approach is not only computationally much more efficient than
neural networks, it also requires much less training data.  As long as
\emph{numerical} trajectory data are used, it is usually fairly cheap to
produce large amounts of data.  However, in principle such approaches can
also be applied to \emph{experimental} data which may be much harder to get
in large quantities.  Of course, this raises then the question how robust
the used methods are against noisy data.  Kernel ridge regression includes
a regularization parameter to mitigate the effects of noise, thus enhancing
the reliability of the discovered conservation laws.

The paper is organized as follows: In Section \ref{II}, we briefly recall
the basic properties of conservation laws.  Section \ref{III} introduces
the concept of an indeterminate regression, the key component of our
method. We explain the process of discovering a single polynomial
conservation law, discuss its implementation and possible validation
strategies, and finally present some examples.  In Section \ref{IV}, we
extend our analysis to the discovery of several conservation laws and
non-polynomial conservation laws.  Section \ref{V} addresses some
complexity considerations.  Section \ref{VI} explores further applications
of our method, including its applicability to discrete dynamical systems
and implicitization problems.  Finally, some conclusions are given.

\section{Conservation laws}\label{II}

In this work, we will be dealing with (continuous) $D$-dimensional
dynamical systems of the form
\begin{equation}\label{eq:dynsys}
    \dot{x}_d = f_d(\mathbf{x})\qquad d=1,\dots,D\
\end{equation}
where the vector field $\mathbf{f}$ on the right hand side is assumed to be
at least once continuously differentiable (most dynamical systems in
applications are at least smooth if not even analytic).  We will usually
assume that \eqref{eq:dynsys} is explicitly given, but for our basic
approach this is not necessary, as it only requires knowledge of finitely
many points on finitely many trajectories.

A \emph{conservation law} or a \emph{conserved quantity} or a \emph{first
  integral} is by definition a function $\Phi\colon\RR^D\to\RR$ which is
constant along each trajectory of the system (\ref{eq:dynsys}).  Assuming
that $\Phi$ is also at least once continuously differentiable and using the
Lie or orbital derivative along the vector field $\mathbf{f}$, this is
equivalent to $\Phi$ satisfying the linear partial differential equation
\begin{equation}\label{eq:pde}
    f_1(\mathbf{x})\frac{\partial\Phi}{\partial x_1} +\cdots+ 
    f_D(\mathbf{x})\frac{\partial\Phi}{\partial x_D} = 0\;.
\end{equation}
As we will be interested only in differentiable conservation laws, equation
(\ref{eq:pde}) provides a simple rigorous validation criterion for
candidate functions $\Phi$ (if the vector field $\mathbf{f}$ is explicitly
given).

Many dynamical systems in physics stem from a variational principle and
their conservation laws are related to variational symmetries via Noether's
theorem \cite{olv:lgde} which often allows their explicit construction with
symmetry methods. By contrast, most biological models are of a more
phenomenological nature without an underlying variational principle and for
them it is much harder to find conservation laws.  While linear ones can
often be easily constructed with linear algebra (see Remark~\ref{rem:plin}
below), nonlinear ones are rarely known. In principle, methods based on
(adjoint) Lie symmetries allow to construct them systematically
\cite{ba:sym}, but for ordinary differential equations it is usually very
hard, if not impossible, to find their Lie symmetries.

A first consequence of the above definition is that any constant function
$\Phi$ defines a conservation law.  Obviously, such conservation laws are
not useful and they are called \emph{trivial}.  In the sequel, we will only
be concerned with non-trivial ones. The definition of a conservation law
also implies that if the system (\ref{eq:dynsys}) admits one, it
automatically admits infinitely many.  Indeed, if $\Phi_{1},\dots,\Phi_{L}$
are some conservation laws, then any function of the form
$g(\Phi_{1},\dots,\Phi_{L})$ for an arbitrary function~$g$ is a
conservation law, too.  If one speaks about finding ``all'' conservation
laws, one actually means finding a maximal set of \emph{functionally
  independent} conservation laws.  Differentiable functions
$\Phi_{1},\dots,\Phi_{L}$ are functionally independent, if and only if
their Jacobian $\partial\boldsymbol{\Phi}/\partial\mathbf{x}$ has almost
everywhere maximal rank meaning that the gradient vectors
$\nabla_{\mathbf{x}}\Phi_{1},\dots,\nabla_{\mathbf{x}}\Phi_{L}$ are almost
everywhere linearly independent.  In physics, a Hamiltonian system of
dimension $D=2d$ admitting $d$ functionally independent conservation laws
is called \emph{completely integrable} \cite{xz:intds}.  A dynamical system
(\ref{eq:dynsys}) may possess up to $D-1$ functionally independent
conservation laws; if this is the case, each trajectory is uniquely
determined by the values of these laws.  A Hamiltonian system with more
than $d$ conservation laws is often called \emph{superintegrable}
\cite{mpw:super}.

\section{Indeterminate Regression}\label{III}
\label{sec:indetreg}

Our approach is based on the literal definition of a conservation law: a
function $\Phi$ which is conserved, i.\,e.\ constant, along trajectories.
Like most approaches presented in the literature, we therefore use
numerical trajectory data and not the dynamical system~(\ref{eq:dynsys})
itself.  Given finitely many points on a trajectory, any conservation
law~$\Phi$ must evaluate to the same value at all these points.  As several
trajectories may lie on the same level set of $\Phi$, we cannot assume that
at points on different trajectories $\Phi$ must evaluate to different
values.  We consider this situation as an \emph{indeterminate regression
  problem}, i.\,e.\ a regression problem where the labels are unknown at
the beginning and must be determined later by additional conditions.

\subsection{Discovering One Polynomial Conservation Law}
\label{sec:poly}

We assume that we are given $N$ points $\mathbf{x}_{m,n}\in\RR^D$ on each
of $M$ trajectories $T_m$ of the dynamical system~(\ref{eq:dynsys}).  One
could easily work with a different number of points on each trajectory.
However, to avoid a bias in the regression, we prefer to take always the
same number of points.  Thus our indices always satisfy $m\in\{1,\dots,M\}$
and $n\in\{1,\dots,N\}$ and we have a total of $MN$ data points.  In
addition, we introduce $M$ yet undetermined labels $y_m$ and our regression
problem consists of finding a function $\Phi$ such that
$\Phi(\mathbf{x}_{m,n})=y_m+\epsilon_{m,n}$ with error terms
$\epsilon_{m,n}$ which are minimal in a suitable sense.  Note that the
desired value for $\Phi(\mathbf{x}_{m,n})$ does not depend on~$n$. This
fact encodes that we search for functions which are constant along
trajectories and hence the number of labels is much smaller than the total
number of data points.  Ha and Jeong \cite{hj:dcl} speak here of ``grouped
data''.

We use \emph{kernel ridge regression}.  Let
$\mathcal{K}\colon\RR^D\times\RR^D\to\RR$ be the chosen kernel.  We will
actually work with the polynomial kernel
$\mathcal{K}_{c,q}(\mathbf{x},\mathbf{x}')=(\mathbf{x}\cdot\mathbf{x}'+c)^q$
parameterised by a non-negative real number $c\geq0$ (set in all our
experiments to $c=1$) and a degree $q\in\NN$.  If $K$ is the corresponding
kernel matrix of dimension $MN$ defined by
$K_{mn,\bar{m}\bar{n}}=\mathcal{K}(\mathbf{x}_{m,n},\mathbf{x}_{\bar{m},\bar{n}})$,
then the coefficients of the unique solution of the regression problem can
be expressed in closed form as
\begin{equation}\label{eq:alpha}
    \boldsymbol{\alpha} = \bigl(K+\lambda \One_{MN}\bigr)^{-1} \mathbf{Y}
\end{equation}
where $\One_{MN}$ denotes the $MN$-dimensional identity matrix, $\lambda$
the regularisation parameter of the ridge regression and
$\mathbf{Y}=\mathbf{1}_N \otimes \mathbf{y}$ the Kronecker product of an
$N$-dimensional vector $\mathbf{1}_N$ consisting only of ones and the
$M$-dimensional label vector $\mathbf{y}$. Here and in the sequel, double
indices $m,n$ are always sorted first by the value of $m$ and then by the
value of $n$.  In other words, we consider first all points on the first
trajectory, then all points on the second trajectory and so on.  The
solution itself is given by the linear combination
\begin{equation}\label{eq:krrsol}
  \Phi(\mathbf{x}) = \sum_{m=1}^M \sum_{n=1}^N
  \alpha_{m,n}\varphi_{m,n}(\mathbf{x})\,.
\end{equation}
of the base functions
$\varphi_{m,n}(\mathbf{x})=\mathcal{K}_{c,q}(\mathbf{x},\mathbf{x}_{m,n})$.

In (\ref{eq:alpha}), it is inconvenient to work with the label
vector~$\mathbf{Y}$, as so many of its entries are identical.  Consider the
matrix $\hat{K}\in\RR^{MN\times M}$ defined as
\begin{equation}
  \hat{K}=\bigl(K+\lambda \One_{MN}\bigr)^{-1}\cdot
  (\One_{M}\otimes \mathbf{1}_N)\,.
\end{equation}
Its first column is the sum of the first $N$ columns of
$\bigl(K+\lambda \One_{MN}\bigr)^{-1}$, its second column the sum of the
next $N$ columns of the inverse and so on.  This allows us to express
(\ref{eq:alpha}) more compactly as $\boldsymbol{\alpha}=\hat{K}\mathbf{y}$.
Defining the vector valued function $\mathbbm{k}(\mathbf{x})$ by
\begin{equation}
    \mathbbm{k}_{\bar{m}}(\mathbf{x}) = \sum_{m=1}^M \sum_{n=1}^N
    \hat{K}_{mn,\bar{m}}\mathcal{K}(\mathbf{x},\mathbf{x}_{m,n})\,,
\end{equation}
we can finally write the solution (\ref{eq:krrsol}) in a form emphasising
the linear dependency on the yet unknown labels~$\mathbf{y}$:
\begin{equation}\label{eq:krrsoly}
  \Phi(\mathbf{x}) =
  \sum_{\bar{m}=1}^M y_{\bar{m}}\mathbbm{k}_{\bar{m}}(\mathbf{x})\,.
\end{equation}

There remains the problem of determining labels~$\mathbf{y}$ such that
\eqref{eq:krrsoly} possibly defines a conservation law.  A simple approach
-- in line with the general philosophy of a ridge regression -- proceeds as
follows: we set one label, say $y_{1}$, equal to $1$ (to avoid the trivial
solution $\mathbf{y}=0$) and determine the remaining labels in $\mathbf{y}$
by the condition that the Euclidean (or $L^{2}$) norm of
$\boldsymbol{\alpha}=\hat{K}\mathbf{y}$ should be minimal.  Obviously, this
condition leads to a quadratic minimisation problem for the labels which
can be easily solved in closed form.  With thus determined labels, we
have finally obtained a candidate $\Phi$ for a conservation law expressed
either in the form \eqref{eq:krrsol} or \eqref{eq:krrsoly}.

\subsection{Practical Realisation}
\label{sec:pract}

Since we are working with ``grouped'' data, one has to choose two
parameters for determining the size of used data set: the number $M$ of
trajectories and the number $N$ of points on each trajectory.  The total
number of points is thus $MN$.  As generically each trajectory provides
information about a further level set of the conserved quantity, we
generally prefer larger values of $M$ and smaller values of~$N$.  For a
reasonable sampling of the phase space, one should probably take $N\geq D$
with $D$ the phase space dimension. Furthermore, the parameters $M$ and $N$
are chosen such that both the hold-out and the training sets in a stratified
five-fold cross-validation can be filled with the same number of points
(thus we typically choose values which are multiples of $25$).

As the points on each trajectory should be about equally spaced to avoid a
sampling bias, we do not integrate the dynamical system \eqref{eq:dynsys}
in the given form, but normalise the vector field $\mathbf{f}$.  Thus we
use as right hand side the field $\mathbf{f}/\|\mathbf{f}\|$.  This
modification does not change the trajectories, but only their time
parametrisation.  If we now take on each trajectory points at times
separated by a fixed time interval $\Delta t$, they are automatically
equally spaced with respect to the arc length.  The initial points for the
different trajectories are randomly picked inside a rectangular cuboid and
we integrate from each initial point both forward and backward.  The size
of the cuboid must be chosen such that the obtained labels do not cluster
too closely around $1$.

Once the data has been produced, the regularisation parameter $\lambda$ and
the labels $\mathbf{y}$ must be determined.  We consider them all as
hyperparameters and compute them simultaneously.  For $\lambda$ we use a
grid search (first on a logarithmic grid for getting the right order of
magnitude and then on a linear grid for refinement) and for $\mathbf{y}$ a
$5$-fold cross-validation.  We first set aside as hold-out set a stratified
random sample of $20\%$ of the generated data points; stratified means here
that we randomly chose on each trajectory $20\%$ of the points on it.

The remaining $80\%$ of the data points are randomly divided in five
disjoint subsets which are again stratified, i.\,e.\ each set contains the
same number of points from each trajectory.  This yields five different
splits where always four of the subsets are used as training set and the
remaining subset as test set.  For each value of $\lambda$ on our grid, we
determine the optimal labels $\mathbf{y}_{\lambda}$ by the condition that
the mean value of the $L^{2}$ norms of the five coefficient vectors
$\boldsymbol{\alpha}_{\lambda}^{(i)}$ obtained for the five different
training sets is minimal.  Since this choice is identical to the loss
function for the individual regressions, there is no trade-off between
$\lambda$ and $\boldsymbol{\alpha}_{\lambda}^{(i)}$.

Each vector $\boldsymbol{\alpha}_{\lambda}^{(i)}$ determines a function
$\Phi_{\lambda}^{(i)}$ which we evaluate on the test points of the $i$th
split.  The result for each point is compared with the entry of the label
vector $\mathbf{y}_{\lambda}$ corresponding to the trajectory on which the
test point lies.  We then compute the root mean squared error over all test
points of all splits and choose that value for $\lambda$ which yields the
minimal error (together with the corresponding label vector
$\mathbf{y}_{\lambda}$).

\begin{figure}
  \centering
  \includegraphics[width=0.9  \linewidth]{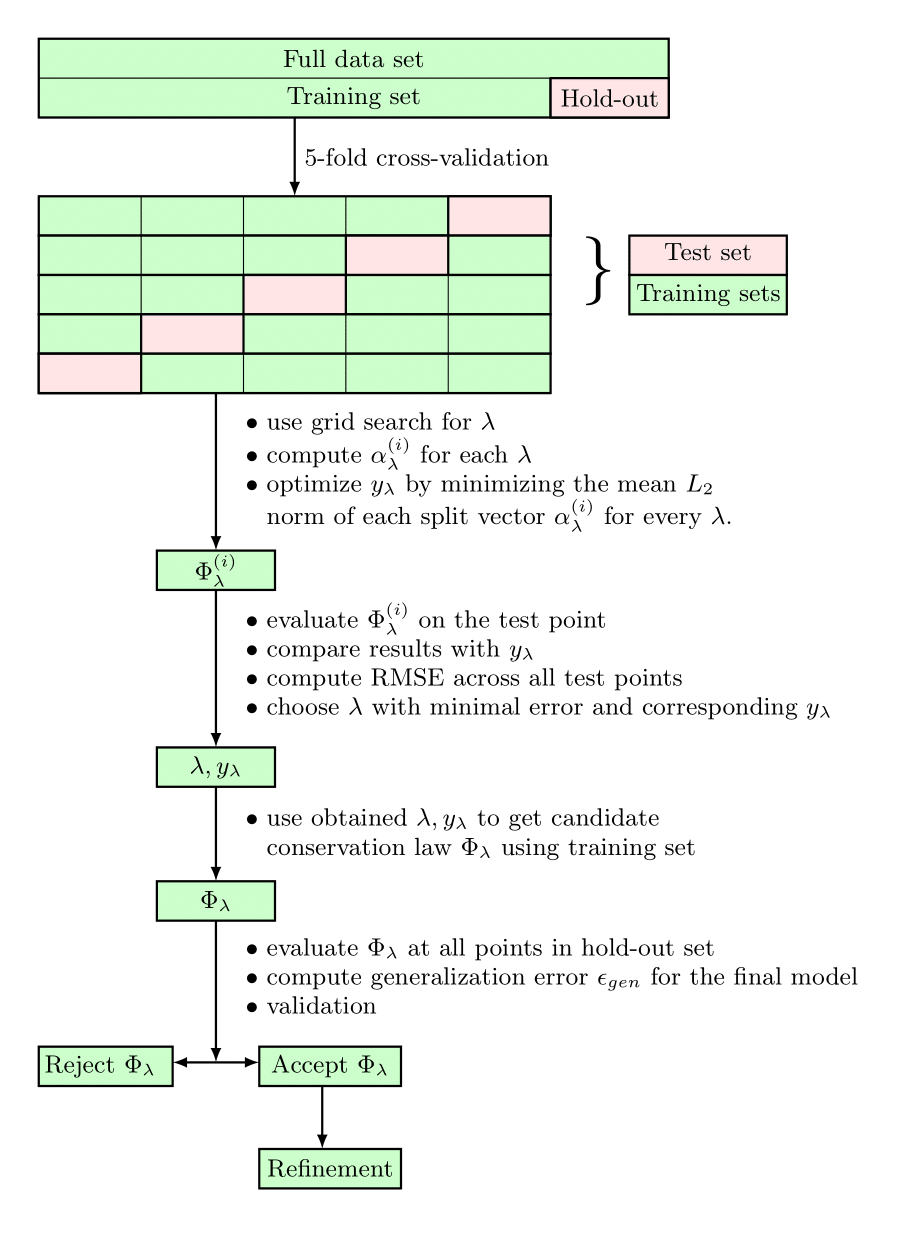}
  \caption{Flowchart of the indeterminate regression.\label{fig:pipe}}
\end{figure}

For the chosen values of $\lambda$ and $\mathbf{y}_{\lambda}$, we determine
the final coefficient vector $\boldsymbol{\alpha}_{\lambda}$ using all data
points outside the hold-out set which then defines our final candidate
conservation law $\Phi_{\lambda}$.  The function $\Phi_{\lambda}$ is then
evaluated at all points in the hold-out set and the results are again
compared with the labels for the corresponding trajectories.  We call the
root mean squared error over all these tests $\epsilon_{\mathrm{gen}}$ and
consider it as a measure for the generalisation error of our final model.
Figure~\ref{fig:pipe} depicts the complete procedure.

Entering the computed vector $\boldsymbol{\alpha}_{\lambda}$ into
\eqref{eq:krrsol} yields our candidate conservation law $\Phi$.  Assuming
that we are using the polynomial kernel~$\mathcal{K}_{c,q}$, this candidate
$\Phi$ is a polynomial -- but written as a linear combination of the base
functions $\varphi_{m,n}(\mathbf{x})$.  As these functions are dense
polynomials and not convenient for a human reader, we explicitly write
$\Phi$ as an expanded polynomial in $D$ variables of degree at most~$q$
\begin{equation}\label{eq:polyPhi}
  \Phi(\mathbf{x})=
  \sum_{0\leq|\boldsymbol{\mu}|\leq q}
  a_{\boldsymbol{\mu}}\mathbf{x}^{\boldsymbol{\mu}}
\end{equation}
where $\boldsymbol{\mu}=(\mu_{1},\dots,\mu_{D})$ denotes an exponent vector
and the $a_{\boldsymbol{\mu}}$ are numerical coefficients determined from
$\boldsymbol{\alpha}$.  For a normalisation, we divide all coefficients by
the one with the largest absolute value, but continue to call them
$a_{\boldsymbol{\mu}}$.

Typically, most of the the obtained coefficients $a_{\boldsymbol{\mu}}$ are
close to zero.  As the largest coefficient is $1$ after the normalisation,
we consider all coefficients with an absolute value several orders of
magnitude smaller as numerical artifacts and set them exactly zero.  In our
experience, a reasonable threshold is often $10^{-5}$ or $10^{-6}$.  As a
``beautification'', one may also think about rounding all coefficients to
the corresponding number of digits, as this often helps to make
coefficients exactly equal which differ only by a very small amount.  But
we will present here in the next section a more rigorous alternative.

\subsection{Validation and Refinement} 
\label{sec:valid}

The above outlined procedure will always produce some candidate function
$\Phi$, but $\Phi$ does not necessarily define a conservation law.  In
fact, the studied dynamical system \eqref{eq:dynsys} might not possess any
conservation law at all -- at least not within the Hilbert space defined by
the used kernel.  Thus we need a validation procedure for accepting or
rejecting a candidate.

There are two fairly immediate possibilities for a purely numerical
validation.  One verifies directly that $\Phi$ is constant along each of
the computed trajectories which can be done independently of the dynamical
system \eqref{eq:dynsys} or one uses the equivalent characterisation of
conservation laws via the partial differential equation \eqref{eq:pde}
which, however, requires explicit knowledge of the vector
field~$\mathbf{f}$.

In the first approach, we compute for each trajectory the mean value
$\bar{\Phi}_{m}$ of the candidate $\Phi$ evaluated at the data points on
this trajectory, i.\,e.\
\begin{equation}\label{eq:mean}
  \bar{\Phi}_{m}=\frac{1}{N}\sum_{n=1}^{N}\Phi(\mathbf{x}_{m,n})
\end{equation}
and then consider the root mean squared \emph{relative} deviation of $\Phi$
from these mean values
\begin{equation}\label{eq:dev}
 \Delta_{1}(\Phi) = \sqrt{\frac{1}{MN}\sum_{m=1}^{M}\sum_{n=1}^{N}
  \frac{\bigl(\Phi(\mathbf{x}_{m,n})-\bar{\Phi}_{m}\bigr)^2}{\bar{\Phi}_{m}^{2}}}
\end{equation}
We accept $\Phi$ as a conservation law, if $\Delta_{1}(\Phi)<\epsilon_{1}$
for a user determined validation threshold~$\epsilon_{1}>0$.

Kernel methods provide us automatically with a candidate $\Phi$ in a
symbolic form without any need for a symbolic regression.  Furthermore, it
is easy to compute its derivatives $\partial\Phi/\partial\mathbf{x}$.
Thus, a direct \emph{symbolic} validation via the partial differential
equation \eqref{eq:pde} seems possible.  However, as the coefficients in
the linear combinations \eqref{eq:krrsol} or \eqref{eq:krrsoly},
respectively, have been determined numerically, we cannot expect that
\eqref{eq:pde} is satisfied exactly.  Thus we resort instead again to a
purely numerical approach by evaluating the left hand side of
\eqref{eq:pde} at all data points $\mathbf{x}_{m,n}$ and considering the
root mean squared error
\begin{equation}\label{eq:devs}
  \Delta_{2}(\Phi)=\sqrt{\frac{1}{MN}
  \sum_{m,n}{\biggl(\sum_{d=1}^{D}f_{d}(\mathbf{x}_{m,n})
    \frac{\partial\Phi}{\partial x_{d}}(\mathbf{x}_{m,n})\biggr)^{2}}}\;.
\end{equation}
We accept $\Phi$ as a conservation law, if $\Delta_{2}(\Phi)<\epsilon_{2}$
for some prescribed threshold~$\epsilon_{2}$.  One should note that while
it is natural to use the given data points for computing
$\Delta_{2}(\Phi)$, one could in principle use any set of test points that
reasonably samples some open subset of the state space.  It is not even
necessary that the points are grouped, i.\,e.\ come from some trajectories.

From a mathematical point of view, the approaches are equivalent and in
experiments one indeed observes a close correlation between them.
$\Delta_{1}(\Phi)$ is easier to compute and as a relative variation easier
to interpret which in turn makes it easier to choose the threshold
$\epsilon_{1}$.  In our experiments, we therefore preferred to work with
$\Delta_{1}(\Phi)$.  

If the vector field $\mathbf{f}$ is explicitly known, one can use the
symbolic evaluation of the partial differential equation \eqref{eq:pde} for
the \emph{refinement} of candidates which have passed the numerical
validation.  Our starting point is the representation \eqref{eq:polyPhi} of
$\Phi$ as an expanded polynomial -- after all very small coefficients have
been set to zero.  Let
\begin{equation}\label{eq:pertPhi}
  \overline{\Phi}(\mathbf{x},\boldsymbol{\delta})=
  \sum_{\substack{0\leq|\boldsymbol{\mu}|\leq q\\ a_{\boldsymbol{\mu}}\neq0}}
  (a_{\boldsymbol{\mu}}+\delta_{\boldsymbol{\mu}})\mathbf{x}^{\boldsymbol{\mu}}
\end{equation}
be a perturbed form of this representation with yet undetermined
perturbations $\delta_{\boldsymbol{\mu}}$.  As typically many coefficients
$a_{\boldsymbol{\mu}}$ in \eqref{eq:polyPhi} will be very small, we can
expect that the number of these perturbations will be much smaller than the
dimension $n_{D,q}=\binom{D+q}{q}$ of the vector space of polynomials of
degree at most $q$ in $D$ variables.

We assume now furthermore that our vector field $\mathbf{f}$ is polynomial,
too, with entries of degree at most $q_{\mathbf{f}}$.  Entering
\eqref{eq:pertPhi} into the left hand side of the partial differential
equation \eqref{eq:pde} and expanding, we obtain a polynomial
\begin{equation}\label{eq:pertval}
  \sum_{d=1}^{D}f_{d}(\mathbf{x})
  \frac{\partial\overline{\Phi}(\mathbf{x},\boldsymbol{\delta})}
  {\partial x_{d}}=
  \sum_{0\leq|\boldsymbol{\nu}|< q+q_{\mathbf{f}}}
  b_{\boldsymbol{\nu}}(\boldsymbol{\delta})\mathbf{x}^{\boldsymbol{\nu}}
\end{equation}
where the coefficients $b_{\boldsymbol{\nu}}(\boldsymbol{\delta})$ depend
linearly on the perturbations $\delta_{\boldsymbol{\mu}}$, since
\eqref{eq:pertPhi} is linear in them and \eqref{eq:pde} is a linear
differential equation.  Any perturbation $\boldsymbol{\delta}$ with
$\Delta_{2}\bigl(\overline{\Phi}(\mathbf{x},\boldsymbol{\delta})\bigr) <
\epsilon_{2}$ is consistent with the used data (obviously, here using
$\Delta_{2}$ is more natural than taking $\Delta_{1}$).

We compute the least-squares solution $\boldsymbol{\delta}^{\ast}$ of the
linear system $b_{\boldsymbol{\nu}}(\boldsymbol{\delta})=0$ for
$0\leq|\boldsymbol{\nu}|<q+q_{\mathbf{f}}$.  Note that despite its outer
appearance this is an inhomogeneous linear system, as the coefficients
$b_{\boldsymbol{\nu}}(\boldsymbol{\delta})$ usually contain constant terms.
Generally, we can expect this system to be overdetermined, as the number of
coefficients $b_{\boldsymbol{\nu}}$ is larger than the number of
perturbations $\delta_{\boldsymbol{\mu}}$ for $q_{\mathbf{f}}>1$ (i.\,e.\ for a
nonlinear dynamical system).  However, depending on the exact form of
$\mathbf{f}$ and $\overline{\Phi}$, it may happen in exceptional cases that
the system is underdetermined.  Then the least-squares solution is not
unique and it is natural to take the unique one of minimal norm.  After
possibly rounding coefficients to a prescribed number of digits, our final
conservation law is then given by
$\overline{\Phi}(\mathbf{x},\boldsymbol{\delta}^{\ast})$.

\subsection{Examples}
\label{sec:exmp}

Most dynamical systems appearing in physics or biology depend on
\emph{parameters} and these also show up in their conservation laws.  We
will therefore always consider the parameters as further state space
variables with trivial dynamics.  Obviously, this approach increases the
dimension $D$ of the dynamical system -- in biological systems often
significantly.  Furthermore, one generally has now to work with a kernel of
higher degree $q$ to accommodate for the dependency on the parameters.

\begin{exmp}
  A classical textbook example of a system with a conservation law is the
  \emph{H\'enon-Heiles system} from astronomy:
  \begin{equation}\label{eq:hh}
    \begin{gathered}
    \dot{q}_{1}=p_{1}\,,\quad  \dot{q}_{2}=p_{2}\,,\\
    \dot{p}_{1}=-q_{1}-2q_{1}q_{2}\,,\quad
    \dot{p}_{2}=-q_{2}-q_{1}^{2}+q_{2}^{2}\,.
    \end{gathered}
  \end{equation}
  It is a Hamiltonian system and thus has as a conserved quantity the total
  energy given by
  \begin{equation}\label{eq:hhE}
    E=\frac{1}{2}(p_{1}^{2}+p_{2}^{2}+q_{1}^{2}+q_{2}^{2}) +
    q_{1}^{2}q_{2}-\frac{1}{3}q_{2}^{3}\,.
  \end{equation}
  The trajectories are rather different for different values of $E$.  For
  $E<1/6$, all trajectories are bounded and regular; for $E>1/6$ most
  trajectories show a chaotic behaviour.

    \begin{figure*}[ht]
    \centering
    \begin{subfigure}{0.26\textwidth}
        \centering
    \includegraphics[width=\textwidth]{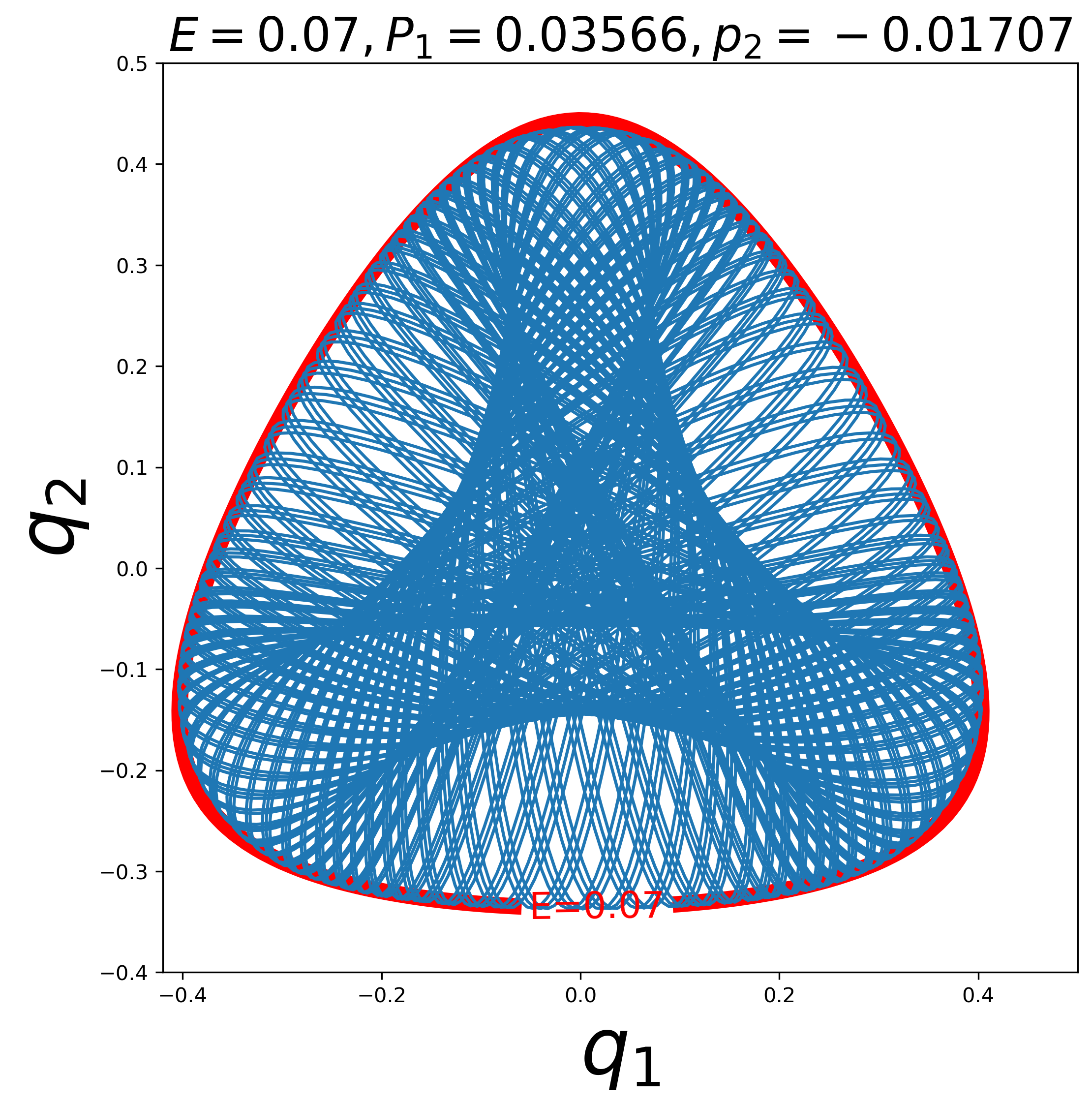}
        \caption{\label{E007}}
    \end{subfigure}
\begin{subfigure}{0.35\textwidth}
        \centering
    \includegraphics[width=\textwidth]{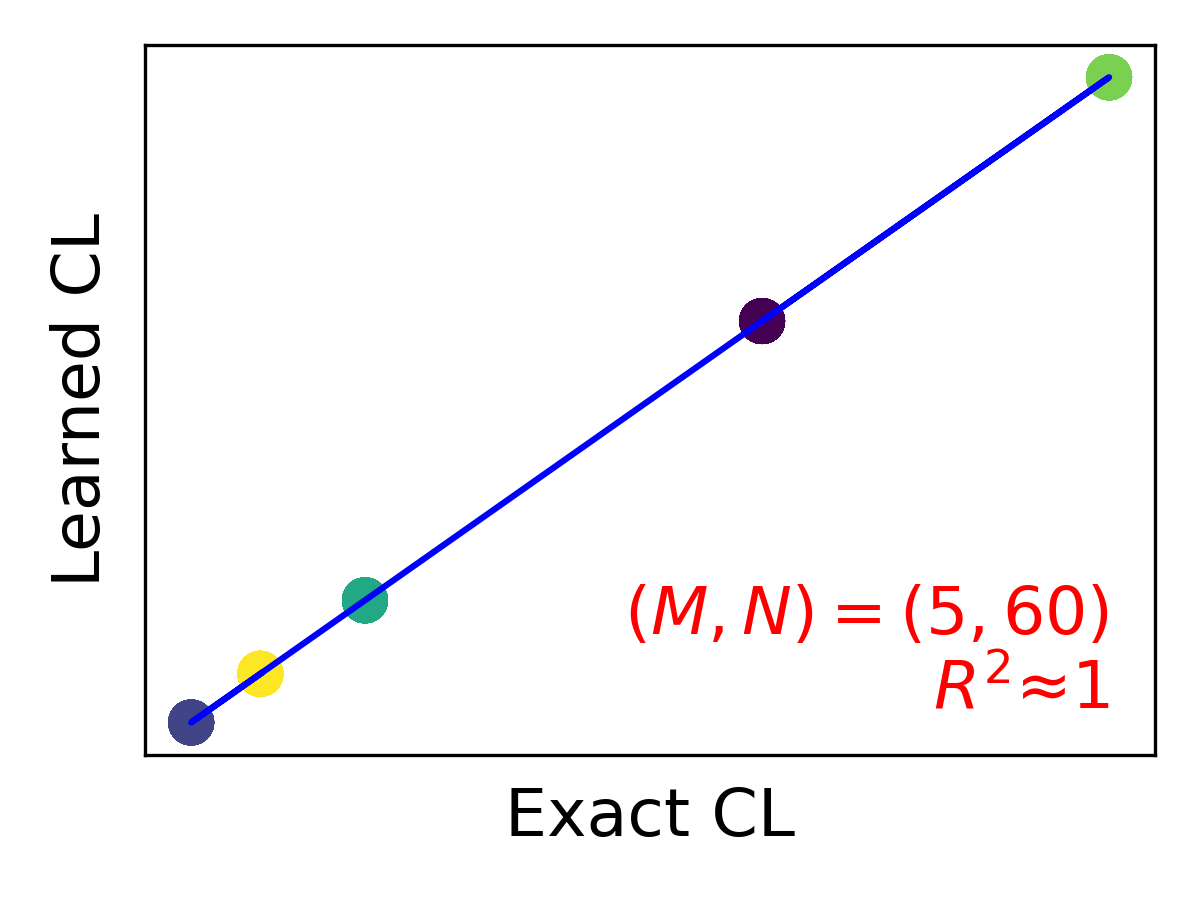}
        \caption{}\label{Exact}
    \end{subfigure}
    \begin{subfigure}{0.35\textwidth}
        \centering
    \includegraphics[width=\textwidth]{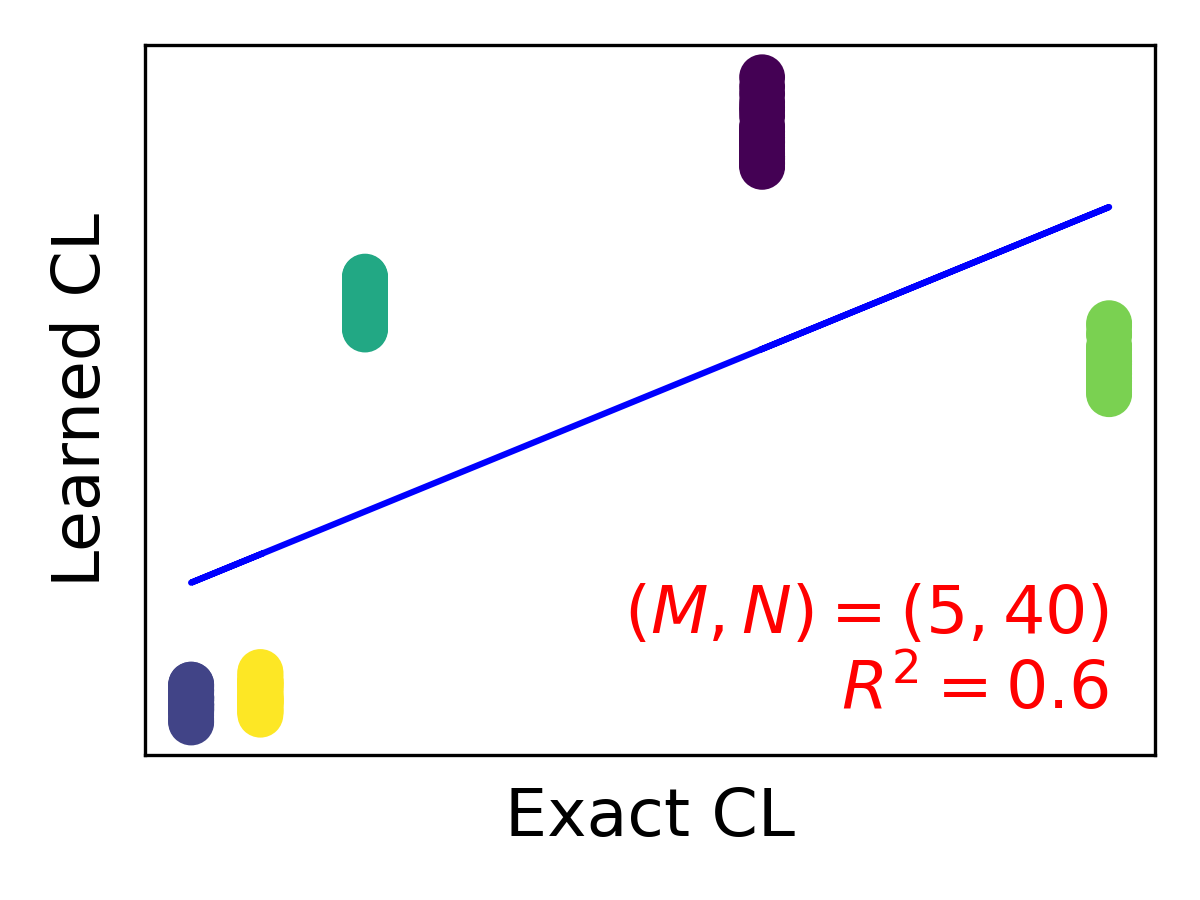}
        \caption{}\label{Exactt}
    \end{subfigure}
    \begin{subfigure}{0.49\textwidth}
        \centering
    \includegraphics[width=\textwidth]{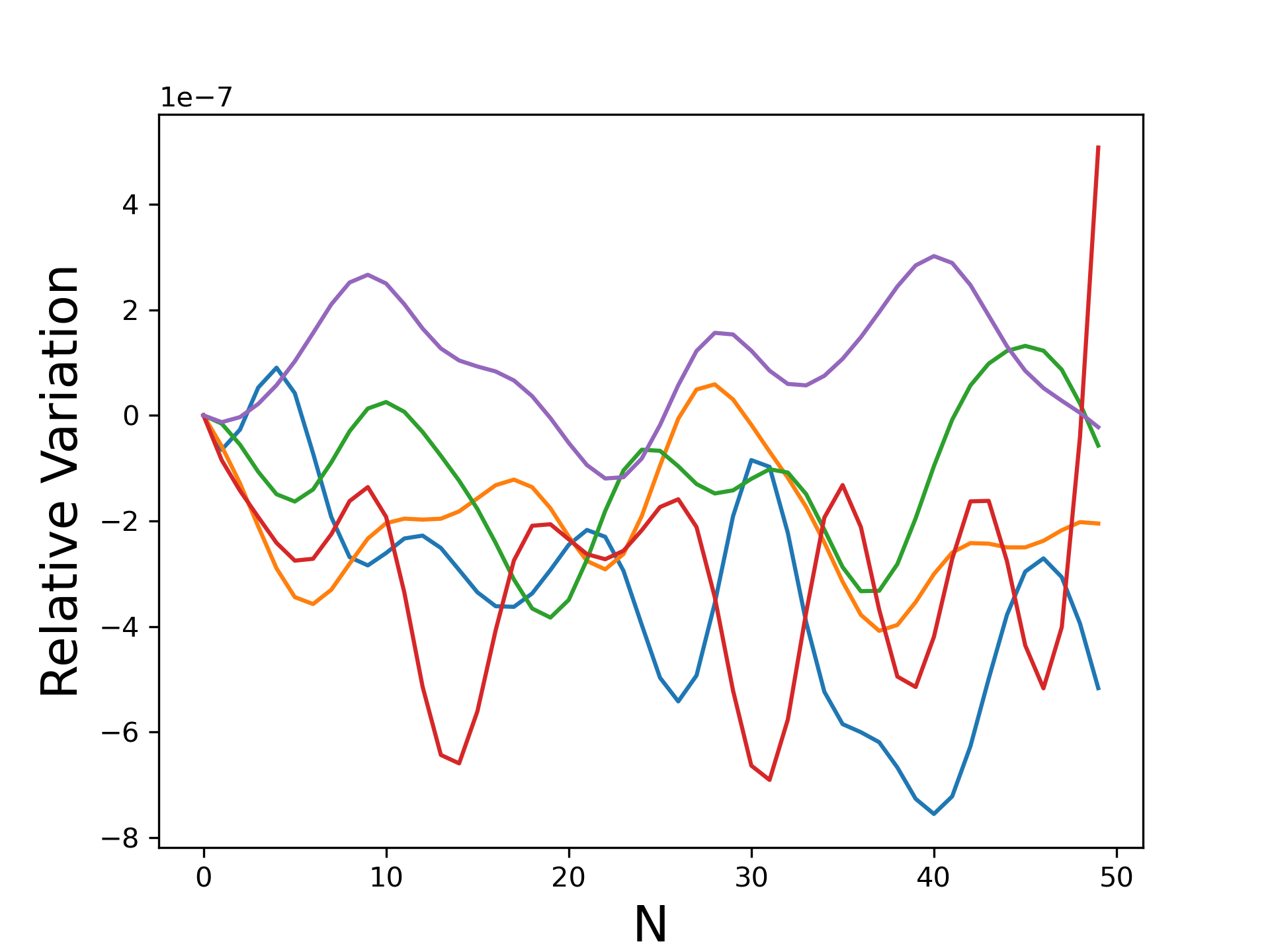}
        \caption{}\label{Hcha}
    \end{subfigure}
    \begin{subfigure}{0.49\textwidth}
        \centering
    \includegraphics[width=\textwidth]{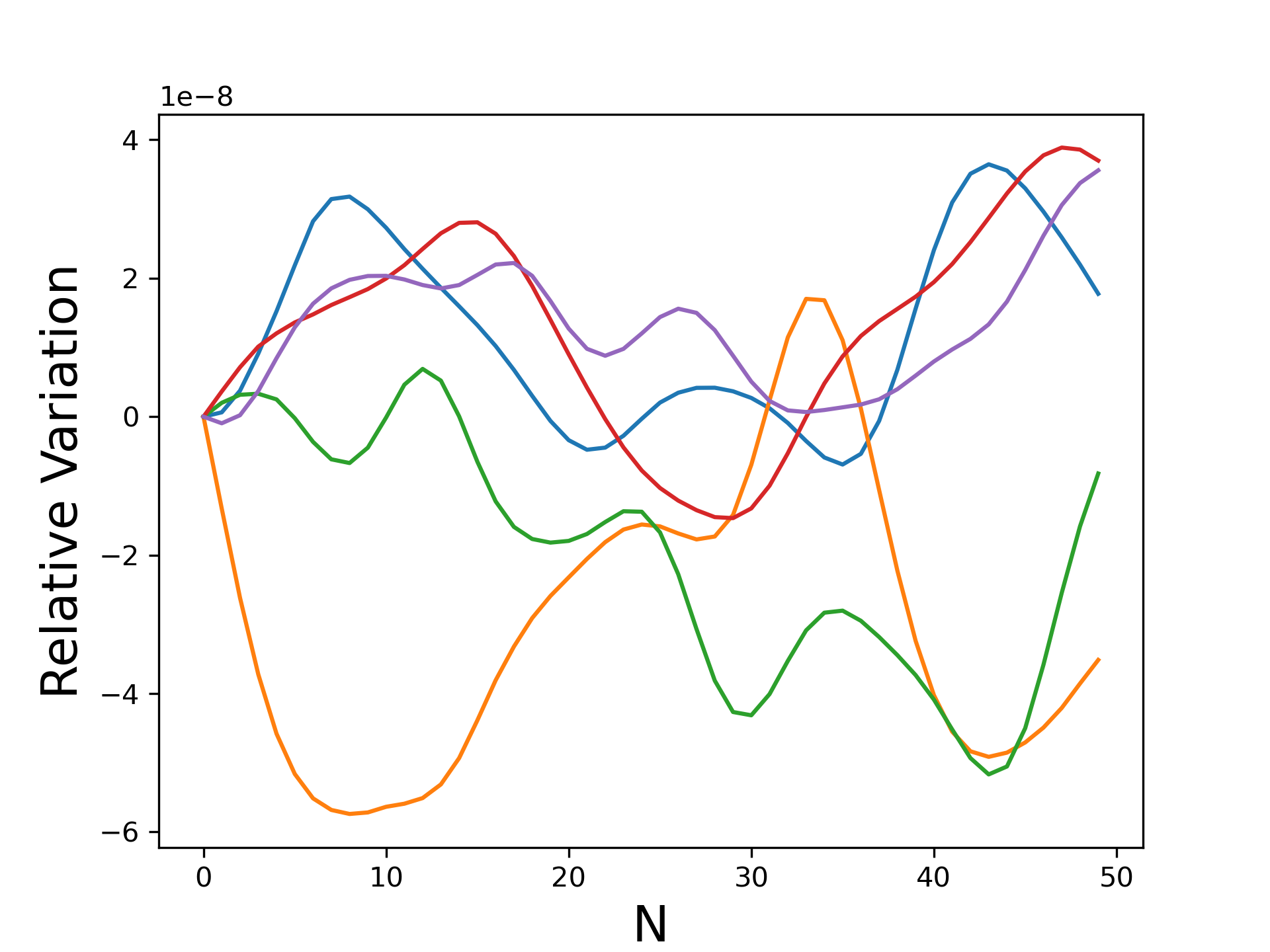}
        \caption{}\label{Hnon}
    \end{subfigure}
    \caption{H\'enon-Heiles model: \eqref{E007} shows the projection to the
      $q_1$-$q_2$-plane of a regular trajectory at energy $E=0.07$.
      Figures \eqref{Exact} and \eqref{Exactt} depict the correlation
      coefficients for 300 and 200 data points, respectively.  The relative
      variations along some random regular and chaotic trajectories are
      shown in \eqref{Hcha} and \eqref{Hnon}, respectively.}
    \label{chanon}
  \end{figure*}

  We conducted some experiments on the H\'enon-Heiles system using
  different initial data sets, some composed entirely of points on chaotic
  trajectories and some using only points on regular trajectories.  Some
  results are shown in Figure~\ref{chanon}.  We were particularly
  interested in estimating how many data points are necessary to discover
  the conservation law using a cubic polynomial kernel.  It turned out that
  with 200 data points no correct polynomial was learned, whereas with 300
  data points good results were achieved.  These numbers should be
  contrasted with the dimension $n_{4,3}=35$ of the space of polynomials of
  degree $3$ in four variables: we needed roughly eight times as many data
  points as coefficients were to be determined.  A further increase of the
  number of data points did not lead to notable improvements in the
  validation.  Our refinement procedure produced here the exact expression
  for the energy eliminating all numerical errors in the coefficients.

  While the basic results were very similar for data sets consisting of
  regular or chaotic trajectories, respectively, one could see some small
  differences in a closer analysis.  It seems that for a fixed total number
  of data points it is better in the chaotic case to use a smaller number
  of trajectories with more points on each of them, whereas in the regular
  case more trajectories with a lower number of points work better.  A
  possible explanation could be that a chaotic trajectory has a higher
  fractal dimension.  On the one hand, it thus contains more information
  about the corresponding level set of the energy, but on the other hand
  more data points are necessary to extract this information.
\end{exmp}

\begin{exmp}\label{ex:fput}
  As a ``scalable'' example, we consider (generalised)
  \emph{Fermi--Pasta--Ulam--Tsingou (FPUT) lattices} \cite{fpu:nlin}
  described by second-order systems of the form
  \begin{equation}\label{eq:fput}
    \ddot{x}_{\ell}=f(x_{\ell+1}-x_{\ell})-f(x_{\ell}-x_{\ell-1})\,.
  \end{equation}
  For simplicity, we restrict to periodic lattices, i.\,e.\ we assume that
  $x_{\ell+L}=x_{\ell}$ for some given $L\in\NN$.  It then suffices to
  consider only $\ell=1,\dots,L$ and rewriting \eqref{eq:fput} as a
  first-order system yields a system of dimension $D=2L$.  The obtained
  system is Hamiltonian so that it possesses a conservation law:
  \begin{equation}\label{eq:fputh}
    H=\frac{1}{2}\sum_{\ell=1}^{L}\dot{x}_{\ell}^{2}+\sum_{\ell=1}^{L}F(x_{\ell+1}-x_{\ell})
  \end{equation}
  where $F$ is an antiderivative of $f$.  One usually assumes that $F$ can
  be represented by a power series and the original FPUT $\alpha$-model is
  obtained for the simplest non-trivial choice
  $F(x)=\frac{1}{2}kx^{2}+\frac{1}{3}\alpha x^{3}$.  In the $\beta$-model,
  one adds one further term $\frac{1}{4}\beta x^{4}$.

  We used the $\alpha$-model treating the parameter $k$ and $\alpha$ as
  further state variables with trivial dynamics.  Hence, our systems were
  of dimension $2L+2$ and we had to use a polynomial kernel of degree $4$.
  In figure \eqref{FPUT}, we show the relative variation along some random
  trajectories for $L=2$ and $L = 5$.  While the results are significantly
  worse for the larger system, they are still fairly good.  In the case
  $L=5$, a set of $200$ data points was not sufficient for discovering the
  conservation law, whereas $500$ data points yielded good results.  If
  this is contrasted with the dimension $n_{12,4}=1820$ of the space of
  polynomials in $12$ variables and maximal degree~$4$, then one sees that
  comparatively very few data points allow for the discovery.

  \begin{figure*}[ht]
    \centering
    \begin{subfigure}{0.43\textwidth}
        \centering
    \includegraphics[width=\textwidth]{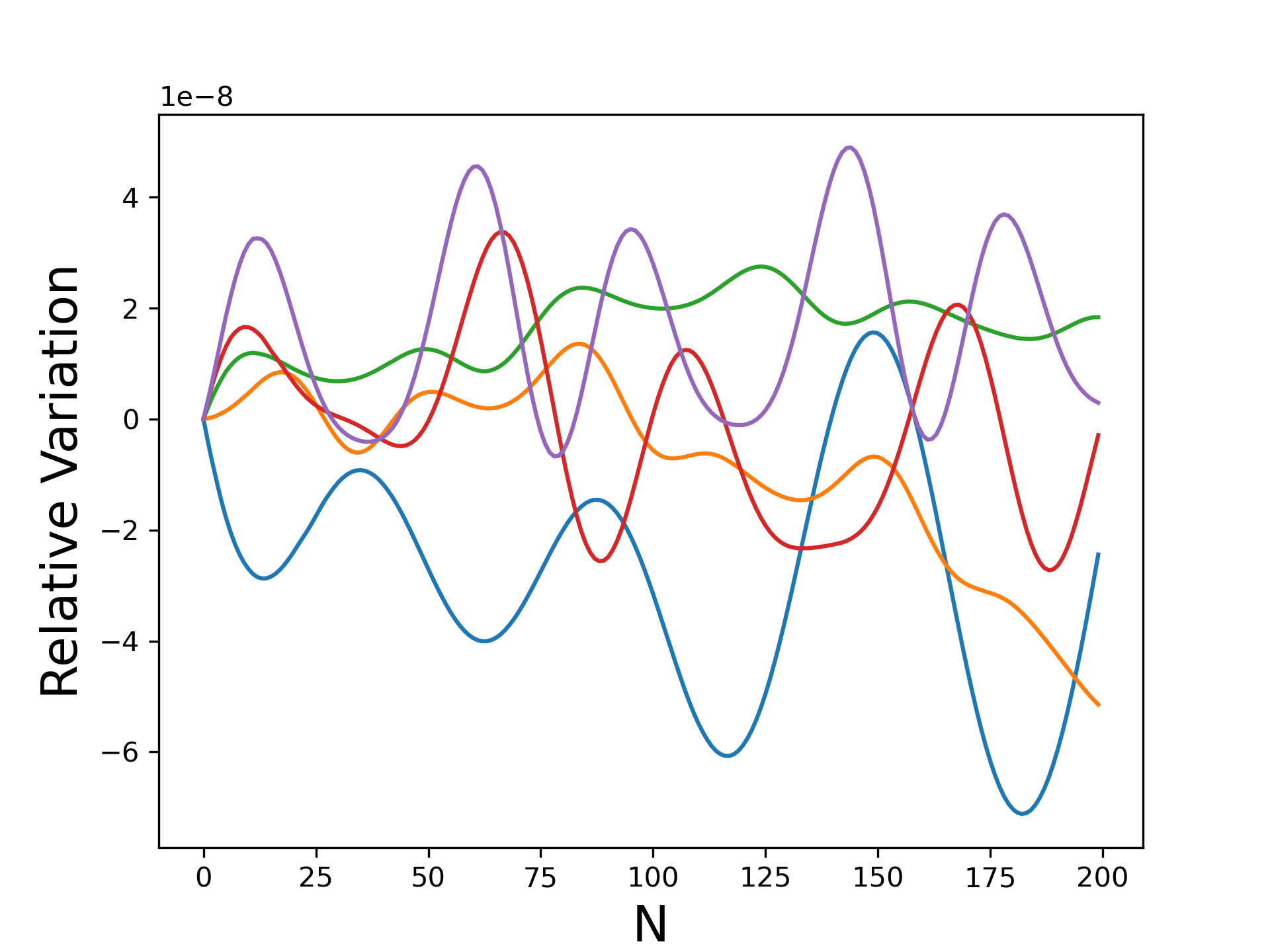}
        \caption{}\label{FPUTn2}
    \end{subfigure}
    \begin{subfigure}{0.52\textwidth}
        \centering    \includegraphics[width=\textwidth]{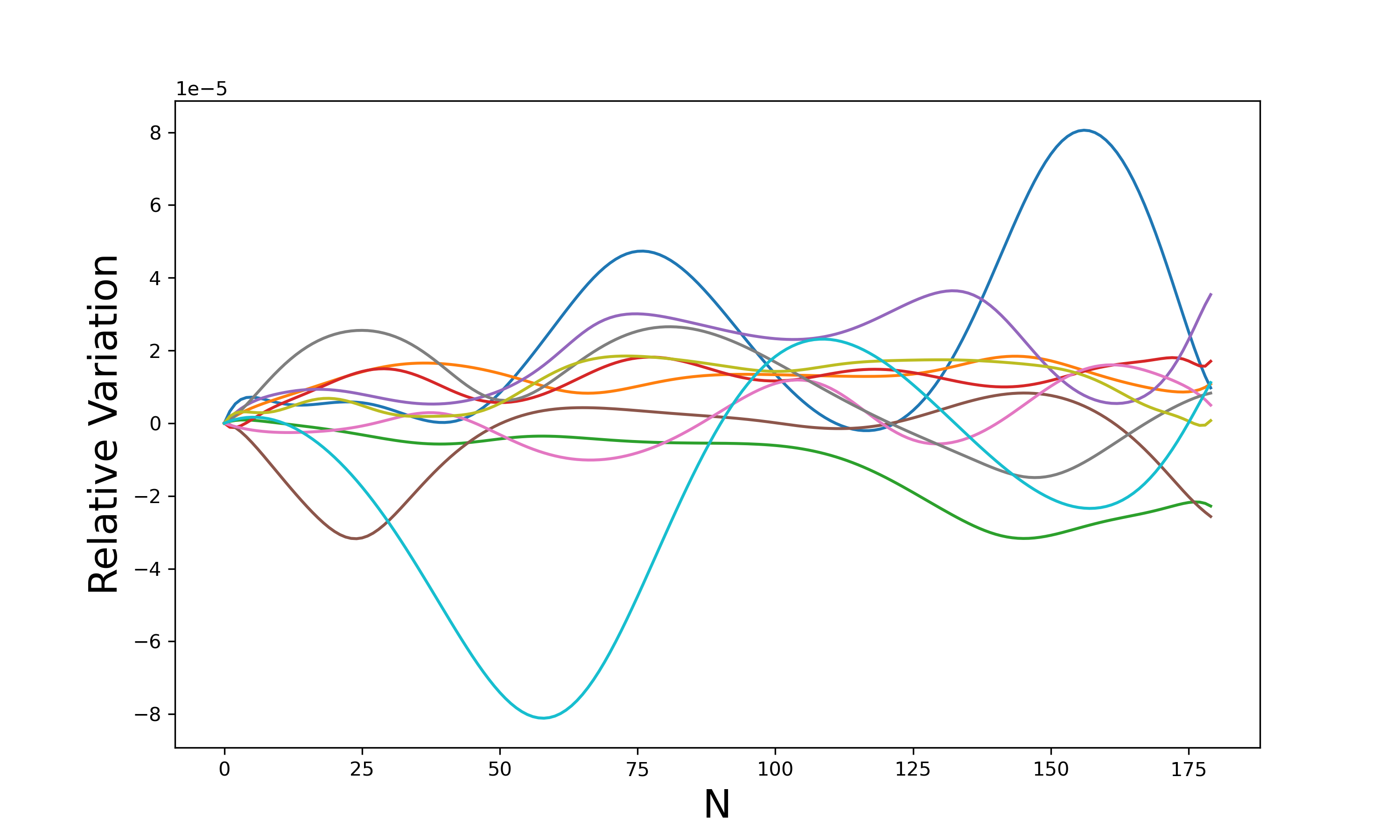}
        \caption{}\label{FPUTn3}
    \end{subfigure}
    \caption{Relative variation of the discovered conservation law along random
      trajectories for FPUT lattices with $L=2$ (left) and $L = 5$ (right) vertices.}
    \label{FPUT}
  \end{figure*}

\end{exmp}

\section{Extensions}\label{IV}
\label{sec:ext}

\subsection{Discovering More Conservation Laws}
\label{sec:more}

A dynamical system might possess several (functionally independent)
conservation laws.  A simple way to discover these consists of an iteration
of the approach presented in Section~\ref{sec:poly} following an idea
already proposed by Wetzel et al.~\cite{wmspg:siam}. It is, however, only
applicable, if it is possible to produce data points for arbitrary initial
conditions, and thus generally cannot be used with experimental data.
Assume that the validation has confirmed the discovery of a first
conservation law~$\Phi_{1}$.  Then we generate new trajectory data such
that all trajectories lie on a level set of $\Phi_{1}$, i.\,e.\ we choose
for the trajectories initial data $\mathbf{x}_{0}^{(i)}$ for $i=1,\dots,M$
such that
$\Phi_{1}(\mathbf{x}_{0}^{(1)})=\cdots=\Phi_{1}(\mathbf{x}_{0}^{(M)})$.
Applying our indeterminate regression to this new data, we will obtain a
new candidate $\Phi_{2}(\mathbf{x})$.  As our ansatz ensures that
$\Phi_{2}$ cannot be constant on all data points, $\Phi_{2}$ must be
functionally independent of $\Phi_{1}$.

If validation confirms $\Phi_{2}$ as a conservation law, we produce again
new trajectory data such that all trajectories lie on common level set of
$\Phi_{1}$ and $\Phi_{2}$.  By the same reasoning as above, the new
candidate $\Phi_{3}$ must be functionally independent of the already found
conservation laws $\Phi_{1}$ and $\Phi_{2}$.  We continue in this manner,
until the validation rejects the last candidate.

Note that this approach does not need any a priori assumption on the actual
number of functionally independent conservation laws.  It automatically
stops, when a complete set has been found \emph{within the considered
  Hilbert space}.  Furthermore, this approach allows for an easy
integration of a priori known conservation laws: one simply uses initial
data such that all known conservation laws are constant on all initial
points.

\begin{rem}\label{rem:plin}
  \emph{Linear} conservation laws are easy to compute directly for almost
  any dynamical systems so that there is no need to resort to machine
  learning for them.  Assume that \eqref{eq:dynsys} can be brought into the
  ``pseudolinear'' form $\dot{\mathbf{x}}=N\mathbf{g}(\mathbf{x})$ with a
  constant matrix $N\in\RR^{D\times B}$ and a $B$-dimensional vector
  $\mathbf{g}$ of ``building blocks''.  Such a structure appears e.\,g.\
  naturally in chemical reaction networks where $N$ is the stoichiometric
  matrix and $\mathbf{g}$ the vector of reaction rates
  \cite{da:crnd,mf:crnt}.  More generally, any polynomial vector field is
  of this form with the vector $\mathbf{g}$ consisting of all terms
  appearing in the vector field.  One can show with elementary linear
  algebra that for any vector $\mathbf{v}\in\ker{N^{T}}$ the linear
  function $\Phi(\mathbf{x})=\mathbf{v}\cdot\mathbf{x}$ defines a
  conservation law and that all linear conservation laws are of this form
  -- see e.\,g.\ \cite{eegssw:hopf}.  Thus, one can precompute a complete
  set of independent linear conservation laws and then use machine learning
  only for discovering additional nonlinear conservation laws functionally
  independent of the linear ones.
\end{rem}
For the generation of the new initial data in each iteration of the above
outline procedure, we use a simple approach.  Assume that we have already
found the conservation laws $\Phi_{1},\dots,\Phi_{k}$.  We randomly choose
a first initial point $\mathbf{x}_{0}^{(1)}$; the values
$c_{i}=\Phi_{i}(\mathbf{x}_{0}^{(1)})$ then define the level set on which
we have to choose further points.  We randomly pick values
$(\bar{x}_{k+1},\dots,\bar{x}_{n})$ and compute one solution of the system
$\Phi_{i}(x_{1},\dots,x_{k},\bar{x}_{k+1},\dots,\bar{x}_{n})=c_{i}$.  If
the (numerical) solution of this system makes problems, we simply discard
our random pick and choose a new one.  We iterate this procedure, until we
have obtained the necessary number of initial points.

Note that it is only for notational convenience that we consider the first
$k$ coordinates $x_{i}$ as unknowns here.  Any choice of $k$ coordinates is
possible here and in practise one should make the choice in dependence of
the actual form of the conservation laws $\Phi_{1},\dots,\Phi_{k}$.  If for
example a conservation law is linear in some variable $x_{i}$, then of
course we simply solve it for this variable and ignore it in the system to
be solved numerically.

\begin{exmp}\label{ex:LVnoln}
  The following three-dimensional \emph{Lotka--Volterra system}
  \cite[Ex.~3]{rs:cllv}
  \begin{equation}\label{eq:LVnoln}
    \dot{x}_{1}=x_{1}(x_{2}-x_{3})\,,\quad
    \dot{x}_{2}=x_{2}(x_{3}-x_{1})\,,\quad
    \dot{x}_{3}=x_{3}(x_{1}-x_{2})
  \end{equation}
  has two polynomial conservation laws of degree~$1$ and $3$, respectively,
  namely
  \begin{equation}\label{eq:LVnolncl}
    \Phi_{1}=x_{1}+x_{2}+x_{3}\,,\qquad \Phi_{2}=x_{1}x_{2}x_{3}\,.
  \end{equation}

  For a linear conservation law, the level sets are hyperplanes so that it
  is trivial to generate initial data on a randomly chosen but fixed level
  set.  When searching for the cubic conservation law, we have here $D=d=3$
  and thus $n_{D,d}=20$.  In our experiments, $50$ data point were not
  sufficient to discover $\Phi_{2}$; we needed $100$.  Figure~\ref{Dr}
  shows again the relative conservation error along random trajectories.
  
  \begin{figure}[ht]
    \centering
    \includegraphics[width=9cm]{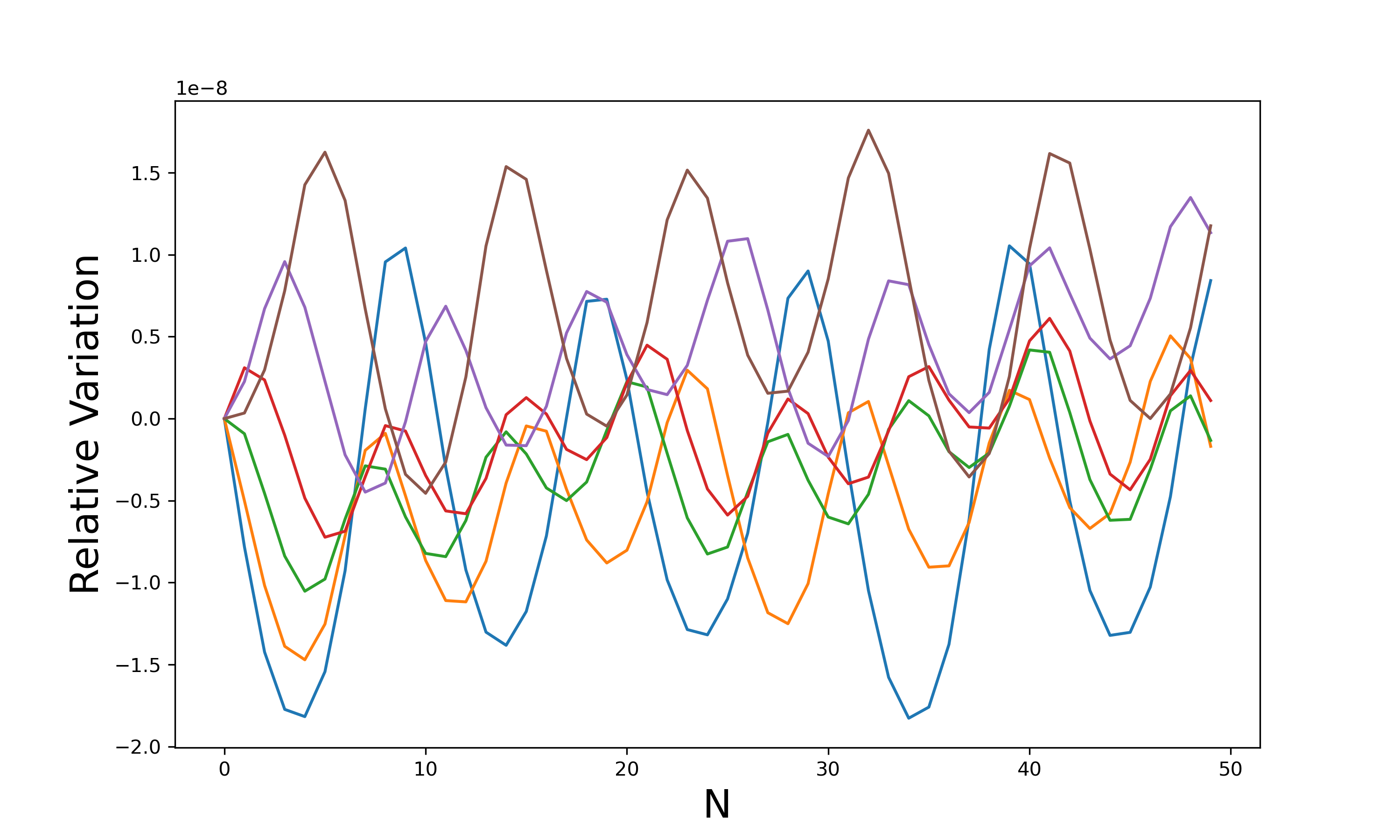}
    \caption{Relative variation of the discovered cubic conservation law
      along random trajectories of the 3D Lotka--Volterra system}
    \label{Dr}
  \end{figure}
\end{exmp}

\begin{exmp}
  The \emph{non-dissipative Lorenz model} is described by the
  three-dimensional system
  \begin{equation}\label{eq:ndlm}
    \dot{x}=\sigma y\,,\quad \dot{y}=-xz+rx\,,\quad \dot{z}=xy
  \end{equation}
  with two parameters $\sigma,r$.  In contrast to the better known chaotic
  Lorenz model, it possesses two conservation laws:
  \begin{equation}\label{eq:ndlmcl}
    \Phi_{1}=\frac{1}{2}x^{2}-\sigma z\,,\quad
    % \Phi_{2}=rx^{2}-\sigma(y^{2}+z^{2})\,.
    \Phi_{2}=rz-\frac{1}{2}y^{2}-\frac{1}{2}z^{2}\,.
  \end{equation}

  A first iteration of our indeterminate regression procedure discovered
  after refinement the exact conservation law $\Phi_{1}$.  Since $\Phi_{1}$
  is linear in $z$ and one considers for $\sigma$ only positive values, we
  produced new initial data with $x,y,\sigma,r$ chosen randomly and $z$
  determined by $\Phi_{1}$.  Due to the parameters, we have here $D=5$ and
  $d=2$ and thus $n_{D,d}=21$.  Already with $40$ data points, $\Phi_{2}$
  is discovered exactly, if one truncates the coefficients to six digits
  (with less data points, it is difficult to perform a cross validation).
  The relative conservation error of the learned $\Phi_{2}$ is shown in
  Figure~\ref{Lorenz}.
  
  \begin{figure}[ht]
    \centering
    \includegraphics[width=9cm]{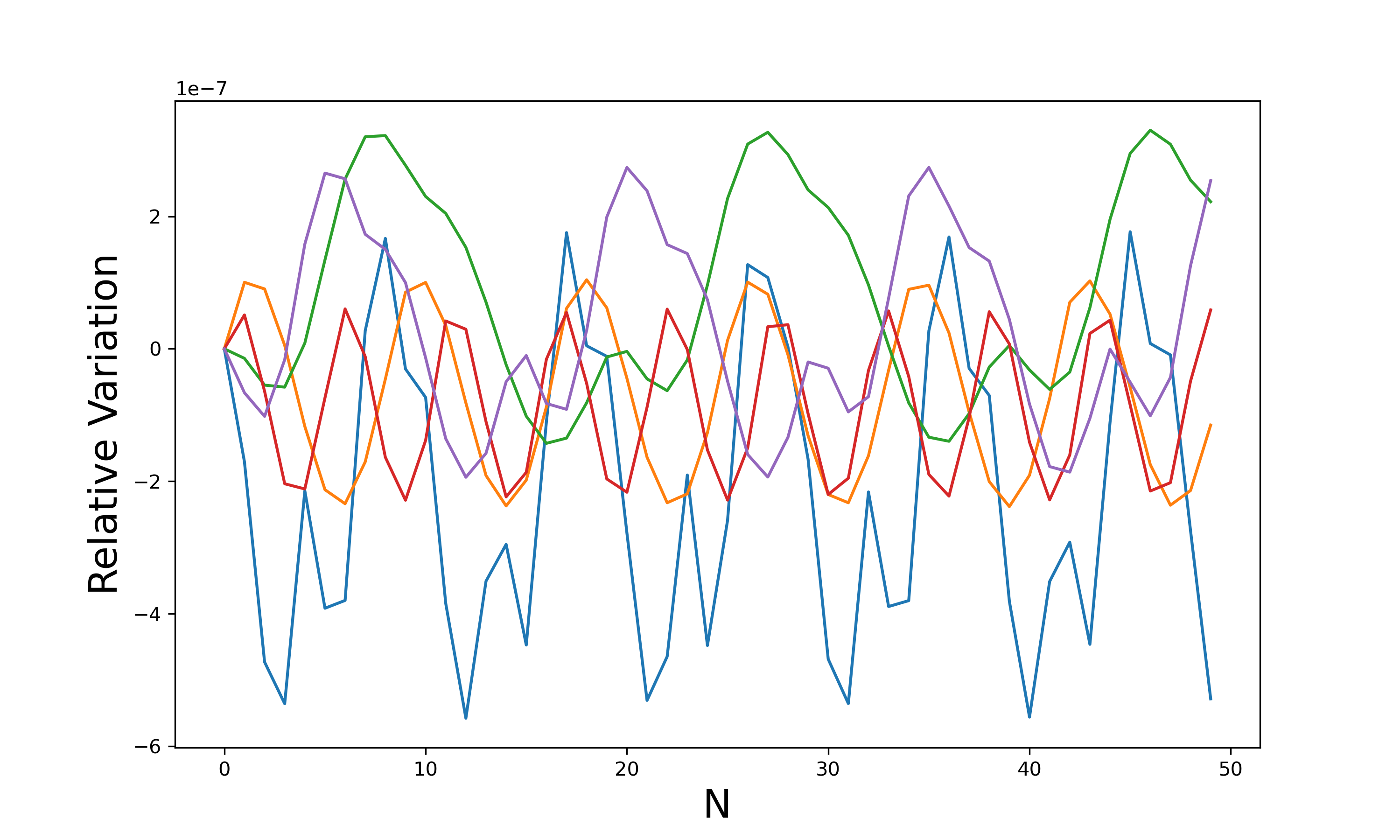}
    \caption{Relative conservation error of $\Phi_{2}$ along random
      trajectories for non-dissipative Lorenz system}
    \label{Lorenz}
  \end{figure}
\end{exmp}

\subsection{More General Conservation Laws}

If $\mathcal{H}$ is the reproducing kernel Hilbert space uniquely
associated with the chosen kernel $\mathcal{K}$, then our approach searches
for conservation laws only in the finite-dimensional subspace of
$\mathcal{H}$ of all functions of the form~\eqref{eq:krrsoly}.  In the case
of an inhomogeneous polynomial kernel of degree~$d$, we can thus only find
conservation laws which are polynomials in $\mathbf{x}$ of maximal
degree~$d$.

If one has a priori knowledge which other, say transcendental or rational,
functions may appear in a conservation law, it is rather easy to adapt our
approach such that also polynomials in these functions are discovered.
Assume we suspect that the functions
$\eta_{1}(\mathbf{x}),\dots,\eta_{L}(\mathbf{x})$ will occur (in
\cite{lsbst:new}, this is called the ``theorist setup'' with known ``basis
functions'').  In our basic approach, the feature vector for the kernel
ridge regression coincides with the coordinate vector~$\mathbf{x}$ of the
dynamical system.  Now, we simply extend the feature vector by the
expressions $\eta_{1}(\mathbf{x}),\dots,\eta_{L}(\mathbf{x})$, before we
start the kernel ridge regression.  Obviously, this will increase the
dimension of the feature vector from $D$ to $D+L$ and thus the
computational costs.  But this approach allows us to find any conservation
law that is polynomial in the variables $\mathbf{x}$ \emph{and} the chosen
transcendental functions $\boldsymbol{\eta}(\mathbf{x})$ (here we have a
slight difference to \cite{lsbst:new} where it is assumed that a
\emph{linear} basis of the considered function space is known).

\begin{exmp}\label{ex:LVln}
  It is well-known that many \emph{Lotka--Volterra systems} admit
  conservation laws containing \emph{logarithmic terms} \cite{rs:cllv}
  (this is also very common in chemical reaction networks, as the entropy
  is a logarithmic quantity \cite{da:crnd}).  This is easily accommodated
  by extending the feature vector: we add for each coordinate $x_{d}$ its
  logarithm $\ln{x_{d}}$, i.\,e.\ we use
  $\bigl(\mathbf{x},\ln{\mathbf{x}}\bigr)$ as feature vector for the
  regression.  For general dynamical systems, one might worry what happens
  for negative~$x_{d}$.  But most biological or chemical models are
  positive and for them only the positive orthant $\mathbf{x}\in\RR^D_{>0}$
  is relevant as state space.

  As a concrete example, we consider the following four-dimensional
  Lotka--Volterra system
  \begin{equation}\label{eq:LVln}
    \begin{gathered}
      \dot{x}_{1}=x_{1}(3-x_{2}-x_{3}-x_{4})\,,\quad
      \dot{x}_{2}=x_{2}(x_{1}-x_{3})\,,\\
      \dot{x}_{3}=x_{3}(-1+x_{1}+x_{2}-x_{4})\,,\quad
      \dot{x}_{4}=x_{4}(-2+x_{1}+x_{3})\,,
    \end{gathered}
  \end{equation}
  belonging to the class considered in \cite[Ex.~7]{rs:cllv}.  It possesses
  the logarithmic conservation law
  \begin{equation}\label{eq:LVlncl}
    \Phi=\sum_{d=1}^{4}(x_{d}-\ln{x_{d}})\;.
  \end{equation}

  As the conservation law $\Phi$ is linear in the variables $\mathbf{x}$
  and $\ln{\mathbf{x}}$, we can work with $d=1$ and $D=8$ and thus have a
  search space of dimension $n_{8,1}=9$.  Nevertheless, we needed $100$
  data points to discover $\Phi$, $50$ points were not sufficient.
  
  \begin{figure}[ht]
    \centering
    \includegraphics[width=9cm]{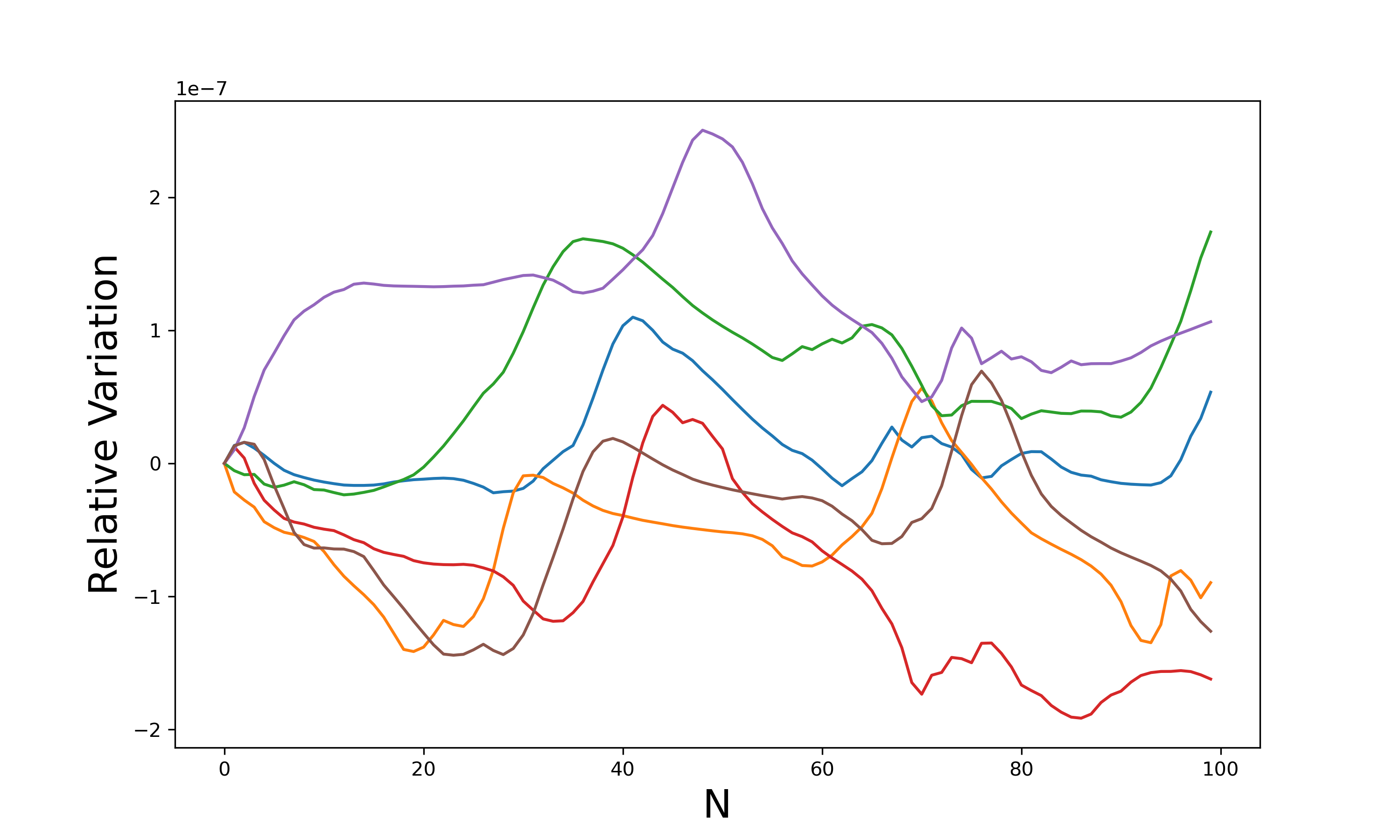}
    \caption{Relative conservation error along random trajectories for 4D
      \emph{Lotka--Volterra} system}
    \label{LV4D}
  \end{figure}
\end{exmp}

\begin{exmp}\label{ex:dsg}
  The \emph{discrete sine-Gordon equation} in the form
  \begin{equation}\label{eq:dsg}
    \ddot{x}_{\ell}=k(x_{\ell+1}-2x_{\ell}+x_{\ell-1})-g\sin{x_{\ell}}
  \end{equation}
  describes a chain of pendula coupled by torsion springs
  \cite[pp.~42--44]{dp:physsolt}, which among other applications is
  sometimes used as a simple model to describe the mechanics of DNA
  \cite{ekhkl:helix}.  For simplicity, we restrict again to periodic chains
  and set $x_{\ell+L}=x_{\ell}$ for some given chain length $L\in\NN$.
  Rewriting \eqref{eq:dsg} as a first-order system yields a system of
  dimension $2L$ which is Hamiltonian with a conservation law
  \begin{equation}\label{eq:dsgh}
    H=\frac{1}{2}\sum_{\ell=1}^{L}\dot{x}_{\ell}^{2}+
    \sum_{\ell=1}^{L}
    \bigl[\frac{k}{2}(x_{\ell+1}-x_{\ell})^{2}+g(1-\cos{x_{\ell}})\bigr]\,.
  \end{equation}

  Pretending that we did not know $H$, we extended the feature vector by
  both $\sin{x_{\ell}}$ and $\cos{x_{\ell}}$ and thus obtained a vector of
  dimension $4L$.  Since we also treated the two parameters $k,g$ as
  dynamical variables, we have here $D=4L+2$ and $d=3$.  We worked with a
  very small chain with $L=3$, so that $n_{D,d}=680$.  With $480$ data
  points we could not discover $H$, but with $600$.  Figure~\ref{DSG} shows
  the relative conservation error along random trajectories.  Again, our
  refinement procedure yielded (after a truncation of the coefficients to
  six digits) the exact conservation law.
  
    \begin{figure}[ht]
      \centering
      \includegraphics[width=9cm]{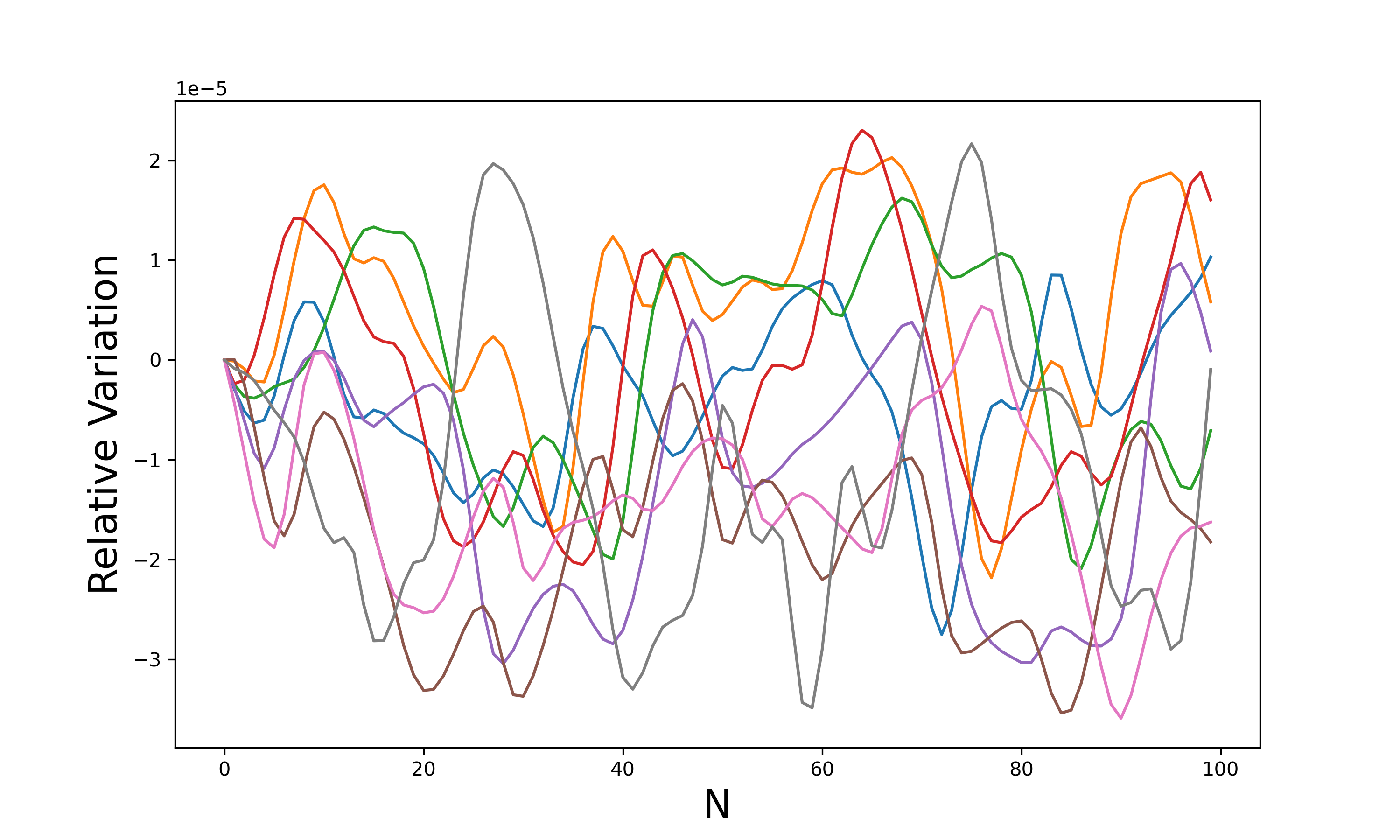}
      \caption{Relative conservation error along random trajectories for
        discrete sine-Gordon equation ($L = 3$)}
      \label{DSG}
    \end{figure}

\end{exmp}

\section{Some Complexity Considerations}\label{V}
\label{sec:complex}

The core computation in our approach is the kernel ridge regression
requiring mainly the inversion of the matrix $K+\lambda\One$ (because of
our indeterminate regression, we indeed need the inverse and not only the
solution of a linear system).  The dimension of this matrix equals the
number $MN$ of used data points.  It is a priori independent of the
dimension $D$ of the feature vector and of the degree~$q$ of the used
polynomial kernel, although our experiments show that typically $MN$ has to
be roughly of magnitude $O(n_{D,q})$ where again $n_{D,q}$ is the dimension
of the space of polynomials in $D$ variables and of degree up to~$q$.  This
implies in particular that modifications of our basic algorithm -- like the
``promotion'' of parameters to dynamical variables or the inclusion of
additional functions into the feature vector $\mathbf{x}$ which all
increase $D$ and may require higher values of $q$ -- have only a moderate
effect on the complexity of this step.  The matrix inversion requires
$O\bigl((MN)^{3}\bigr)$ operations.  The fact that our matrix is symmetric
and positive definite leads to better constants in this estimate, but does
not change the cubic complexity.  The (system) dimension $D$ is relevant,
when the data is produced by numerical integration.  One may estimate the
costs for this step to be roughly $O(DMN)$ which is, however, neglectable
compared to the costs of the matrix inversion.

As soon as the final coefficient vector $\boldsymbol{\alpha}$ has been
computed and the candidate conservation law $\Phi$ defined by
\eqref{eq:krrsol} has been validated, we are in principle done and have
found a conservation law.  Compared with approaches where conservation laws
are represented by neural networks, our result is much more explicit, as it
is a closed form expression (with numerical coefficients).  However, as the
conservation law is given as a linear combination of the base functions
$\varphi_{m,n}$, it does not arise in a form convenient for any subsequent
analysis.

One should therefore transform the conservation law $\Phi$ obtained in the
form \eqref{eq:krrsol} into the usual representation \eqref{eq:polyPhi} of
a polynomial.  One may consider this rewriting as the analogue in our
method to the symbolic regression step used in approaches based on neural
networks to obtain a closed form expression for the discovered conservation
law.  If one does the expansion straightforwardly in a symbolic
computation, then each of the base functions
$\varphi_{m,n}(\mathbf{x})=(\mathbf{x}^{T}\mathbf{x}_{m,n}+c)^{q}$ expands
into a dense polynomial of degree $q$ in $D$ variables and we must take a
linear combination of $MN$ such polynomials.  Any such polynomial contains
$n_{D,q}$ terms.  Indeed, we noticed in experiments that -- already for
moderate values of $D$ and $q$ -- such a symbolic computation leads to a
bottleneck.

However, it is straightforward to compute directly the coefficients of the
representation \eqref{eq:polyPhi} in a purely numerical manner via a
multinomial expansion.  We first introduce the following notations.  Let
$\hat{\boldsymbol{\mu}}= (\mu_{0},\mu_{1},\dots,\mu_{D})=
(\mu_{0},\boldsymbol{\mu})$ be an exponent vector with $D+1$ entries and
$\boldsymbol{\mu}$ the vector with $D$ entries obtained by removing
$\mu_{0}$.  We can then write the base functions in the form
\begin{equation}\label{eq:multinomial}
  \varphi_{m,n}(\mathbf{x})=
  \sum_{|\hat{\boldsymbol{\mu}}|=q}\binom{q}{\hat{\boldsymbol{\mu}}}
  c^{\mu_{0}}\mathbf{x}_{m,n}^{\boldsymbol{\mu}}\mathbf{x}^{\boldsymbol{\mu}}
\end{equation}
with $\tbinom{q}{\hat{\boldsymbol{\mu}}}$ a multinomial coefficient and
our conservation law expands to
\begin{equation}
  \Phi(\mathbf{x})
  =\sum_{m=1}^{M}\sum_{n=1}^{N}\sum_{|\hat{\boldsymbol{\mu}}|=q}
  \alpha_{m,n}\binom{q}{\hat{\boldsymbol{\mu}}}
  c^{\mu_{0}}\mathbf{x}_{m,n}^{\boldsymbol{\mu}}\mathbf{x}^{\boldsymbol{\mu}}\;.
\end{equation}
By simply collecting the terms with the same power product, we obtain for
the coefficients $a_{\boldsymbol{\mu}}$ in \eqref{eq:polyPhi} the formula
\begin{equation}\label{eq:amu}
  a_{\boldsymbol{\mu}}= c^{\mu_{0}}\binom{q}{\hat{\boldsymbol{\mu}}}
  \sum_{m=1}^{M}\sum_{n=1}^{N}
  \alpha_{m,n}\mathbf{x}_{m,n}^{\boldsymbol{\mu}}\;,
\end{equation}
where $\mu_{0}=q-|\boldsymbol{\mu}|$.  The numerical evaluation of this sum
is straightforward.  The total cost of the rewriting is then $O(n_{D,q}MN)$
and neglectable against the cost of the kernel ridge regression.

One may wonder why we bother with machine learning techniques, if we search
only for polynomial conservation laws -- in particular, in view of our
refinement method.  One could instead directly make a simple ansatz for a
general polynomial of degree $q$ in $D$ variables with indeterminate
coefficients, enter the ansatz into the partial differential equation
\eqref{eq:pde} and then solve the arising linear system of dimension for
its $n_{D,q}$ coefficients.

For an exact computation, one must then assume that the right hand side of
the dynamical system \eqref{eq:dynsys} consists at most of rational
functions over the rational numbers~$\QQ$, as only then entering a
polynomial ansatz yields after clearing denominators a linear system
over~$\QQ$.  The usual complexity estimates for nullspace computations
assumes that any arithmetical operation can be performed in constant time.
While this is true for floating point operations, it is not true for
computations over the rationals~$\QQ$.  A naive implementation of Gaussian
implementation will even have an exponential bit complexity, as the number
of digits will double in each elimination step.  With more advanced methods
like modular arithmetics, one can again obtain algorithms with a cubic
complexity.  However, the constants are much worse than in a floating point
computation and depend on the size of the results.  As a rule of thumb, one
can treat on a single processor machine linear systems over $\QQ$ up to a
size of about a few thousand indeterminates within reasonable computation
times.

If one replaces the exact nullspace computation of the linear system by a
numerical one, one has a complexity of $O(n_{D,q}^{3})$.  Assuming that
$MN$ and $n_{D_{q}}$ are of the same order of magnitude (which is roughly
the case in all examples that we studied), we find the same complexity as
in our approach.  A key difference is that in our approach it is not
necessary to know the dynamical system \eqref{eq:dynsys} explicitly; it
suffices, if enough trajectory data are available.  If \eqref{eq:dynsys} is
explicitly given, then our refinement step takes into account only those
power products which according to the regression analysis actually appear
in the conservation law.  Thus generally in our refinement step one has to
solve a much smaller linear system and its costs are again neglectable.

In principle, one can also work with an ansatz containing more general
functions than polynomials.  The ``theorist'' approach of \cite{lsbst:new}
is an example how this can be realized purely numerically leading to a
linear regression problem.  For exact computations, their approach is not
suitable, as it would lead to a linear system over a function field instead
of the rationals~$\QQ$.  For doing exact linear algebra computations over
such a field, one must be able to decide whether or not an expression is
zero.  Richardson's theorem (see \cite[Sect.~5]{bl:as}) asserts that this
problem becomes rapidly undecidable, if one adds transcendental functions
like $\exp{x}$ or $\sin{x}$.

However, with a suitable setup we can always reduce to linear algebra over
the rationals.  We first choose a (finite-dimensional) $\QQ$-linear
function space $V_{1}$ which represents the search space in which we look
for conservation laws.  Then we have to construct a second
(finite-dimensional) $\QQ$-linear function space $V_{2}$ which must contain
for each function $\Phi\in V_{1}$ and for each variable $x_{d}$ the product
$(\partial\Phi/\partial x_{d}) f_{d}$.  Let $\{g_{1},\dots,g_{R}\}$ be a
linear basis of $V_{1}$ and $\{h_{1},\dots,h_{S}\}$ one of $V_{2}$.  Making
the ansatz $\Phi=\sum_{r=1}^{R}\beta_{r}g_{r}$ with yet undetermined
coefficients $\beta_{r}\in\QQ$ and entering it into the partial
differential equation \eqref{eq:pde}, we obtain a condition of the form
$\sum_{s=1}^{S}c_{s}(\boldsymbol{\beta})h_{s}=0$ where the coefficients
$c_{s}$ are linear in the unknowns~$\boldsymbol{\beta}$.  The ansatz
defines an exact conservation law, if and only if
$c_{s}(\boldsymbol{\beta})=0$ for $s=1,\dots,S$.  This condition defines a
linear system over $\QQ$ for the coefficients~$\boldsymbol{\beta}$.
Compared with the purely polynomial case, the dimensions $R$ and $S$ are
rapidly growing here for more complicated situations, so that exact
computations will often be unfeasable.

\section{Further Applications}\label{VI}

The basic idea of our approach is fairly independent of the notion of a
conservation law of a dynamical system.  It is concerned with predicting
functions from knowing points in their level sets (the ``grouped data'' of
Ha and Jeong \cite{hj:dcl}).  But such questions appear also in other
situations.  We briefly discuss here two applications.  The first one is
very close to what we have done so far: we consider \emph{discrete}
dynamical systems which can be handled by our approach without any real
changes.  The second one is quite different: the implicitisation of curves
or higher-dimensional surfaces which is important in algebraic geometry,
geometric modeling and computer vision.

\subsection{Discrete Dynamical Systems}

Our approach can also be applied to \emph{discrete} dynamical systems,
i.\,e.\ systems where the time $t$ is a discrete variable.  We assume
$t\in\ZZ$ and consider an autonomous first-order system of the form
\begin{equation}\label{eq:discrsys}
  \mathbf{x}_{t+1}=\mathbf{f}(\mathbf{x}_{t})\,.
\end{equation}
A \emph{conservation law} of it is a function $\Phi\colon\RR^D\to\RR$ which
remains constant along solutions of \eqref{eq:discrsys}, i.\,e.\ which
satisfies $\Phi(\mathbf{x}_{t+1})=\Phi(\mathbf{x}_{t})$ (see e.\,g.\
\cite{ph:sfidiff}).

Thus we are in exactly the same situation, as in the continuous case: we
know points on level sets of $\Phi$ and can apply the approach developed
above.  One may even say that the discrete case is slightly easier to
handle, as evaluation of \eqref{eq:discrsys} along a trajectory does not
require the use of numerical approximations as in the case of differential
equations.  On the other hand, it is more difficult to provide a more or
less uniform sampling of the state space, as one has no control about the
distance between consecutive points.

\begin{rem}
  In the differential case, it was not really necessary to assume that we
  treat a first-order system.  Most numerical methods are geared towards
  such systems, but e.\,g.\ there also exist methods for second-order
  systems.  For our approach this is irrelevant; we only need the
  trajectory data (this is different for the approach by Liu et
  al.~\cite{lmt:poin2} using the partial differential
  equation~\eqref{eq:pde} which assumes a first-order system).  In the
  discrete case, the above definition of a conservation law is valid only
  for first-order systems.  For a system of order~$Q$, conservation laws are
  functions defined on $\RR^{QD}$, as they depend not only on
  $\mathbf{x}_{t}$, but also on the points
  $\mathbf{x}_{t+1},\dots,\mathbf{x}_{t+Q-1}$, i.\,e.\ on a whole segment
  of the trajectory~\cite{ph:sfidiff}.  In principle, our approach can be
  adapted to such a situation, but it is probably easier to rewrite the
  given dynamical system as a first-order one.
\end{rem}

\begin{exmp}\label{ex:discr}
  We consider the second-order difference equation
  \begin{equation}\label{eq:hydon2}
    x_{t+2}=\frac{t}{t+1}x_{t}+\frac{1}{x_{t+1}}
  \end{equation}
  appearing in \cite{ph:sfidiff}.  One readily verifies that
  \begin{equation}\label{eq:hydonCL}
    \Phi(t,x_{t},x_{t+1})=t x_{t} x_{t+1} - \frac{1}{2} t (t+1)
  \end{equation}
  is constant along solutions and thus defines a polynomial conservation
  law.  For applying our approach to this equation, we first rewrite it as
  a first-order system by introducing $y_{t}=x_{t+1}$ and then render it
  autonomous by introducing $z_{t}=t$.  This yields the rational system
  \begin{equation}\label{eq:hydon1}
    x_{t+1}=y_{t}\,,\quad
    y_{t+1}=\frac{x_{t}z_{t}}{z_{t}+1} + \frac{1}{y_{t}}\,,\quad
    z_{t+1}=z_{t}+1
  \end{equation}
  with a polynomial conservation law of degree $3$
  \begin{equation}\label{eq:hydonCL1}
    \Phi(x_{t},y_{t},z_{t})=x_{t}y_{t}z_{t}-\frac{1}{2}z_{t}(z_{t}+1)\,.
  \end{equation}
  Applying our approach to \eqref{eq:hydon1} and rounding the obtained
  coefficients to four digits, we discover the exact conservation law
  \eqref{eq:hydonCL1}.
\end{exmp}

\begin{rem}
  As the approach of Arora et al.~\cite{abbh:conslaw} is based on finding
  via discrete gradients a particularly well adapted discretisation of the
  given continuous dynamical system, it cannot be extended to discrete
  dynamical systems.  Similarly, the approach of Kaiser et
  al.~\cite{kkb:dcl} uses the Koopman theory of continuous dynamical
  systems and cannot be extended.  The neural deflation approach of Zhu et
  al.~\cite{zzk:mlclnd} only applies to Hamiltonian systems and requires a
  Poisson structure.  While there exist discrete analogues of these
  concepts, it is unclear whether an extension is possible.  The approach
  of Liu et al.~\cite{lmt:poin2} is based on learning solutions of the
  partial differential equation~\eqref{eq:pde}.  For discrete dynamical
  systems, one can derive a discrete analogon to this partial differential
  equation where partial derivatives are replaced by shift operators
  \cite{ph:sfidiff}.  It is unclear how solutions of this difference
  equation can be learned, but our refinement procedure could probably be
  adapted to this partial difference equation.  By contrast, the approaches
  of Ha and Jeong \cite{hj:dcl} via a noise-variance loss and Wetzel et
  al.~\cite{wmspg:siam} via Siamese neural networks, respectively, can also
  be straightforwardly extended to the discrete case.
\end{rem}

\subsection{Implicitisation of Curves and Surfaces}

In geometry, two different types of representations of objects like curves
and surfaces (of dimension $2$ or higher) in some affine space $\RR^{D}$
are typically used.  In an \emph{explicit} representation, a
$K$-dimensional object is given via a parametrisation
$x_{d}=\varphi_{d}(t_{1},\dots,t_{K})$ with parameters
$(t_{1},\dots,t_{K})\in P\subseteq\RR^{K}$ from some real parameter set.
In an \emph{implicit} representation, the object is described as the zero
set of some functions $\Phi_{j}\colon\RR^{D}\to\RR$.  Each representation
is preferable for certain tasks.  For example, an explicit representation
allows us to produce easily points on the object, whereas an implicit
representation is better for checking whether a given point lies on the
object.  It is therefore of importance to be able to switch between these
two types of representations and implicitisation is the problem of going
from a given parametrisation to an implicit representation.  An extensive
discussion in the context of geometric modeling can be found in
\cite{cmh:geosolmod}.

In algebraic geometry, the given parametrisation $\varphi_{d}$ is typically
rational and one looks for polynomials $\Phi_{j}$, i.\,e.\ it is assumed
that the represented object is a variety.  Then implicitisation can be
formulated as an elimination problem and techniques like resultants or
Gr\"obner bases are available for its solution (see \cite{cmh:geosolmod}
and references therein).  However, these techniques are expensive and often
lead to polynomials of high degree making subsequent computations costly
and numerically unstable (which is a problem for applications in CAGD).

Consequently, numerous methods for approximate implicitisation have been
developed (see the references in the recent work \cite{ggp:aai}) and our
approach is related to a class of methods called \emph{discrete approximate
  implicitisation}.  The authors of \cite{wlzscyn:encX} propose to use an
autoencoder for fitting polynomial equations, i.\,e.\ a neural network.
Like for conservation laws, we consider our approach via kernel methods as
much cheaper.

While many of the algebraic methods can -- at least in principle -- handle
arbitrary dimensions $K<D$, most of the approximate approaches concentrate
on the case $K=D-1$ for $D=2,3$, i.\,e.\ on planar curves and surfaces in
3D, as these are the dominant situations in CAGD and geometric modelling.
This restriction means that the implicit representation consists of a
single polynomial.  For simplicity, we also consider here only this case.

Our discovery of conservation laws of ordinary differential equations from
trajectory data corresponds to the implicitisation of whole \emph{families}
of curves: we are looking for an implicit representation valid not only for
a single curve, but for many curves simultaneously.  This means that we
search for functions $\boldsymbol{\Phi}$ such that the $k$th curve is
described by the equations $\boldsymbol{\Phi}(\mathbf{x})=\mathbf{c}^{(k)}$
for some constant vector $\mathbf{c}^{(k)}$.  The extension of our approach
to families of two or higher dimensional surfaces is straightforward: one
only needs a way to generate sufficiently many data points on the surfaces
that do not lie on lower-dimensional subsets.  While for a single curve or
surface always an implicitisation exists (though not necessarily one with
polynomials as desired in algebraic geometry), this represents a special
property for families.

In the case of a single curve or surface, one needs in addition to data
points on the curve or surface one point \emph{off} the curve or surface;
this point should not lie too close.  Then we take as labels for
all points on the curve or surface $0$ and for the one additional point
$1$.

\begin{exmp}
  A classical planar algebraic curve is the \emph{trifolium} shown in
  Figure~\ref{fig:trifolium}.  A trigonometric parametrisation of it is given by
  \begin{equation}\label{eq:paratrif}
    \begin{aligned}
      x&=4\sin{(t)}^{4}-3\sin{(t)}^{2}\,,\\
      y&=-\sin{(t)}\cos{(t)}(4\sin{(t)}^{2}-3)
    \end{aligned}
  \end{equation}
  with $0\leq t\leq \pi$.  As an implicit description of it, one may use
  the polynomial equation of degree~$4$
  \begin{equation}
    \label{eq:impltrif}
    (x^{2}+y^{2})^{2}=x(x^{2}-3y^{2})\,.
  \end{equation}

  \begin{figure}[ht]
    \centering
    \includegraphics[width=5cm]{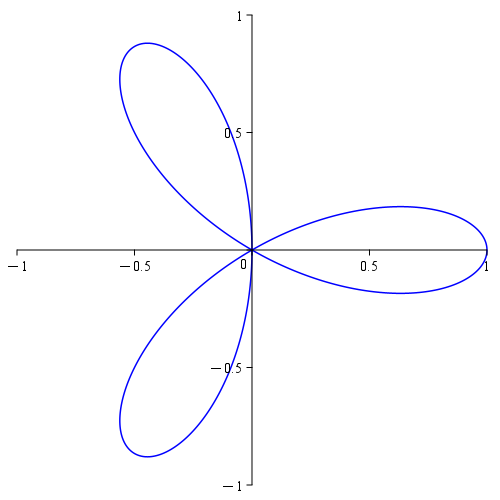}
    \caption{Regular trifolium}
    \label{fig:trifolium}
  \end{figure}

  A point outside of the curve is for example $(-1,0)$.  By employing our
  kernel method, we discover -- after rounding the coefficients to five
  digits -- the exact solution
  $x^4 - x^3 + 2x^2y^2 + 3xy^2 + y^4= (x^{2}+y^{2})^{2}-x(x^{2}-3y^{2})$.
\end{exmp}

\begin{exmp}
  \emph{Whitney's umbrella} is the algebraic surface shown in
  Figure~\ref{fig:whitney} -- note that the ``handle'', i.\,e.\ the
  $z$-axis, is part of the surface.  It is implicitly defined by the
  polynomial equation of degree~$3$
  \begin{equation}\label{eq:implwhit}
    x^{2}z=y^{2}\;.
  \end{equation}
  A polynomial parametrisation of the two-dimensional part of the surface
  is given by
  \begin{equation}
    \label{eq:parawhit}
    x=u\,,\quad y=uv\,,\quad z=v^{2}\,.
  \end{equation}

  \begin{figure}[ht]
    \centering
    \includegraphics[width=5cm]{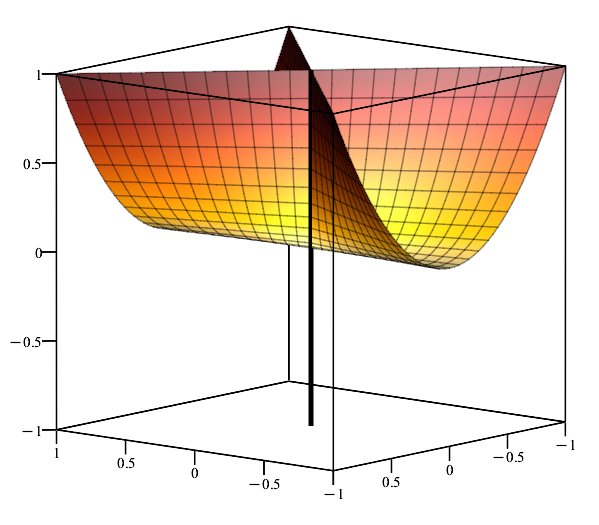}
    \caption{Whitney's umbrella}
    \label{fig:whitney}
  \end{figure}
  
  Points on the $z$-axis are not really necessary for the implicitisation;
  it suffices to consider only points on the two-dimensional part of the
  surface.  A point obviously off Whitney's umbrella is $(1,1,-1)$.  By
  employing a polynomial kernel of degree~$3$, we discover after rounding
  the coefficients to $7$ digits the exact implicit description.
\end{exmp}

\section{Conclusions}\label{VII}

In this work, we proposed an approach to discover conservation laws of
dynamical systems which is based on a kernel method, namely an
``indeterminate'' ridge regression, instead of some variants of neural
networks.  We believe that this is computationally more efficient and in
particular requires much less data which is important, if the data comes
from experiments and not from numerical simulations.  Indeed, in all our
examples a few dozens to a few hundred data points were sufficient ($600$
points was the maximum we used).  By contrast, in
\cite{abbh:conslaw,wmspg:siam,zzk:mlclnd}, the authors report that for
problems of similar sizes they trained their neural networks with $50,000$
to $200,000$ points; only the $2,000$ data points used in \cite{hj:dcl} are
somewhat close to our values.  The neural networks typically consist of
about $200$ to $2,000$ neurons in the hidden layers and the training
requires between $1,000$ and $50,000$ epochs.  The computational costs of
such trainings are probably several orders of magnitudes larger than the
matrix inversion needed in a kernel ridge regression, even if one takes
into account that the determination of the labels in our indeterminate
regression and the tuning of the regularisation parameter require repeated
inversions.  The comparatively low number of data points ensures that the
costs for the inversions remain moderate.  We also emphasize that our
approach yields immediately a symbolic representation of a discovered
conservation law, whereas methods based on neural networks still need a
subsequent symbolic regression step which usually also comes with
considerable computational costs.

The underlying idea of our approach is more general than the problem of
discovering conservation laws: it is about learning functions from
information about their level sets and thus should be adaptable to other
problems, too.  We presented here only a basic version of our approach and
applied it to a number of popular example systems.  There are many
questions that call for a deeper analysis.  An obvious one is the
robustness with respect to noisy data which is again very important, if one
thinks of using experimental data.  One could hope that up to a certain
noise level the regularization parameter of the kernel ridge regression may
compensate the noise.  But only experiments will tell whether this is
really the case.

Our current approach to discover multiple conservation laws requires to
produce new data in each step adapted to the already found conserved
quantities.  Thus it cannot be applied with experimental data.  In
principle, it is not difficult to formulate optimization problems that
iteratively discover functionally independent conservation laws using
always the same data, e.\,g.\ following ideas presented in \cite{lmt:poin2}
or \cite{zzk:mlclnd}.  However, these problems are no longer simple ridge
regressions where one has a closed-form expression for the exact solution,
but must be solved numerically.  Thus one has to study more closely the
properties of these problems and available solution methods.

Finally, we want to mention that our approach should also be able to
discover \emph{approximate} conservation laws.  These could be for example
quantities that are almost conserved on shorter time scales, but show some
evolution on longer time scales or quantities that remain conserved only
after an initial transient phase.  While in physics many models are
constructed in such a way to possess exact symmetries and thus related
exact conservation laws, we expect that in the phenomenological models that
prevail in biology such approximate conservation laws are much more
frequent than exact ones.  In our approach, one can study such phenomena by
a suitable selection of the used data.  Thus one could e.\,g.\ consider
only trajectory points up to a certain maximal time or conversely only
points at times larger than a certain minimal time.  By playing with the
time threshold, one may even glean information about the corresponding time
scales.

\section*{Data Availability}

A repository with the Python code and the data used for the examples
presented in this article is publicly available on Zenodo at the DOI
\url{https://doi.org/10.5281/zenodo.11279856}.

\begin{acknowledgments}
  The work of MAM and WMS was performed within the Research Training Group
  \emph{Biological Clocks on Multiple Time Scales} (GRK 2749) at Kassel
  University funded by the German Research Foundation (DFG).
\end{acknowledgments}

\bibliography{ConsLaw}

\end{document}